\documentclass[floatfix,pra,%
 amsmath,amssymb,
 aps,
longbibliography
]{revtex4-2}

\usepackage{graphicx}
\usepackage{dcolumn}
\usepackage{bm}
\usepackage{nomencl}
\makenomenclature

\usepackage[utf8]{inputenc}
\usepackage[T1]{fontenc}
\usepackage{xcolor}

\newcommand{\rev}[1]{\textcolor{black}{#1}}

\begin{document}

\preprint{APS/123-QED}

\title{Architecting mechanisms of damage in topological metamaterials}

\author{Leo~de Waal}
 \email{ldewaal@ed.ac.uk}
\author{Matthaios~Chouzouris}%
\author{Marcelo~A.~Dias}%
\email{mdias@ed.ac.uk}
 \affiliation{Institute for Infrastructure and Environment, School of Engineering, University of Edinburgh}

\date{\today}

\begin{abstract}
Architecting mechanisms of damage in metamaterials by leveraging lattice topology and geometry poses a vital yet complex challenge, essential for engineering desirable mechanical responses. Of these metamaterials, Maxwell lattices, which are on the verge of mechanical stability, offer significant potential for advanced functionality. By leveraging their robust topological features, they enable precise control of effective elastic properties, manipulation of stress localisation and delocalisation across specific domains, and targeted global damage that follows local fracture events. In this work, we identify topology and geometry-dependent parameters that establish a simple, yet precise, framework for designing the behaviour of non-idealised Maxwell lattices and their damage processes. We numerically explore the underlying phenomenology to demonstrate how this framework can guide or arrest damage in lattices, both with and without domain walls and additional boundary constraints. Our approach uncovers a robust way to manipulate the mechanisms of damage and the path they follow in metamaterials, with further insight into crack arrest, diversion, and shielding.
\end{abstract}

\maketitle


\section{Introduction}

Significant efforts have been made to determine the relationship between fracture toughness and the underlying microstructure of lattice-based materials~\citep{gibson2003cellular,chen1998fracture,fleck2007damage,tankasala20152013,ATHANASIADIS2021101411,BerkacheKamel2022Meot,omidi2023fracture,choukir2023role}. Beyond metrics related to the onset of failure, the resistance to crack growth during the damage propagation has also been explored~\citep{schmidt2001ductile,tankasala2020crack,hsieh2020versatile,wang2024superior}, highlighting an opportunity to increase their steady-state fracture toughness. Harnessing the change in failure characteristics of materials straddling the rigidity transition~\citep{driscoll2016role}, where the fracture delocalises, may also provide a pathway to increase this toughening behaviour. Attempts to design the damage evolution have generally focused on the inclusion of heterogeneities to control the percolating pathway through: material property~\citep{gao2020crack} or density~\citep{manno2019engineering} variations, anisotropic unit cell design~\citep{domino2024fracture, gao2024damage}, and disorder~\citep{karapiperis2023prediction}. The focus on leveraging heterogeneities to influence damage mechanisms have over-passed the rich design space within their homogeneous counterparts. As a result, limited resources are available to accurately predict the stress profile and damage evolution in an already damaged homogeneous lattice-based material, in terms of simplified intuitive principles, derived from the underlying microstructure.

A general indication of the dominant behaviour of a lattice is given by its \textit{coordination number}, $z$. In \textit{d} dimensions, lattices with $z<2d$ ($z>2d$), are bending (stretching)-dominated~\citep{guest2003determinacy}. 
\rev{Maxwell Lattices are on the verge of mechanical stability, thus containing an equal number of nontrivial zero energy modes, or \emph{floppy modes} (FM), and \emph{states of self-stress} (SSS). By bridging the gap between Maxwell Lattices and the topological band theory of electronic systems, it has been shown that the localisation of these surface FM and SSS can be manipulated through a topological polarisation $\textbf{P}_\mathrm{T}$~\citep{kane2014topological}.}
The robustness of this topological quantity provides a powerful tool that can be leveraged to architect the deformation and stress localisation in these lattices~\citep{paulose2015selective,driscoll2016role,chapuis2022mechanical}.

Building upon concepts derived from the robustness of topological properties, efforts to control the mechanics of damage processes in Maxwell lattices have yielded significant results, mainly through the introduction of SSS achieved by combining topologically distinct domains~\citep{zhang2018fracturing,liu2023stress,widstrand2023stress,widstrand2024robustness}. For instance, in idealised (zero-thickness hinges) lattices, it was shown through numerical simulations that SSS domain walls can be used to focus the stress and prevent failure from initiating at imperfections \citep{zhang2018fracturing}. However, when these lattices are manufactured in practice, the joints will inevitably exhibit a non-zero bending rigidity (non-idealised). This was shown to dilute the stress focusing effect~\citep{liu2023stress,widstrand2023stress,widstrand2024robustness} through the delocalisation of the stress associated with \emph{next nearest neighbour} (NNN) coupling, analogous to phenomena described in the case of brittle materials~\citep{salman2021localizing}. By considering NNN interactions, it has been illustrated how the unit-cell geometry can be tuned to adjust this delocalisation~\citep{liu2023stress}, although the damage evolution has not yet been investigated. Furthermore, while the inclusion of NNN provides insights into the delocalisation effects, further discretisation is required to accurately capture the behaviour and damage propagation, particularly at larger values of the unit-cell size to strut thickness, \emph{i.e.} \emph{slenderness ratio} (SR), where bending energy becomes an important contributor.

As damage propagates through a lattice, constraints are removed as the individual elements fracture, necessarily resulting in the removal of SSS and an emergence of FM, thus establishing a certain compliance in the system. Understanding how to control the form of these additional modes opens up a vast design space that amends to the role of topology. Although the edge localization is topologically robust, there is great latent potential for adjusting the form of these modes in lattices that fall under the same equivalence class, which is defined uniquely through the direction of the polarisation vector $\textbf{P}_\mathrm{T}$. This, in turn, provides a yet unrealised tool for precisely controlling the stress concentration and thus a mechanism that can guide the damage path, without necessarily altering $\textbf{P}_\mathrm{T}$. Furthermore, while it has been shown that SSS domain walls can focus stress~\citep{zhang2018fracturing, liu2023stress, widstrand2023stress, widstrand2024robustness}, to the best of our knowledge no insights have yet been provided on how FM, due to imperfections, influence the development and damage of lattices with or without these SSS domain walls.

In this work, we demonstrate how topological techniques can be applied in conjunction with more precise geometrical considerations to guide the damage and mechanical response for non-idealised lattices. A numerical simulator with beam elements is built to capture the fracture propagation and we systematically probe the underlying phenomenology to elucidate how the idealised FM, arising at the crack edge, influence the stress and damage in these lattices. Using this understanding, we then show how topological domain walls, added to the lattice, arrest and control the damage. Finally, we probe the effect of the FM counterpart, the SSS, added to the lattice through additional boundary constraints, and illustrate how these dominate the lattices stress profile and damage process. In addition to advancing our understanding of damage processes in metamaterials, these results are fundamental for paving the way towards architecting fracture paths themselves, providing a framework that can be leveraged to protect, divert, and arrest cracks that occur in lattices.

\section{Background Theory}\label{sec2}
\subsection{The Role of Topology}

The global count of FM ($N_\mathrm{0}$) and SSS ($N_\mathrm{s}$) in a lattice are quantities determined by the topology of the network itself, and are captured by the Maxwell-Calladine index theorem~\citep{maxwell1864calculation,calladine1978buckminster}, which relates them to the number of sites ($N$) and constraints ($N_\mathrm{c}$) in the system:
\begin{equation}\label{Eq:1} 
N_\mathrm{0}-N_\mathrm{s}=Nd-N_\mathrm{c}.
\end{equation} Beyond their global count, however, the above relation does not capture any information regarding the form and localisation of these modes. 

To that end, a kinematic analysis of the unit-cell can be used to connect the site displacements, $\mathbf{u}$, to the bond elongations, $\mathbf{e}$, through the \textit{compatibility matrix} $\mathbf{C}$ (\emph{i.e.} $\mathbf{e}=\mathbf{C}\cdot\mathbf{u}$). The FM can then be found in the nullspace of $\mathbf{C}$, signifying combinations of node displacements that do not stretch any bonds. Likewise, the site forces, $\mathbf{f}$, can be related to the bond tensions, $\mathbf{t}$, using the \textit{equilibrium matrix} $\mathbf{Q}$ (\emph{i.e.} $\mathbf{f}=\mathbf{Q}\cdot\mathbf{t}$). Therefore, the nullspace of $\mathbf{Q}$ contains the SSS. Invoking the principle of virtual work, it can be shown that $\mathbf{C} = \mathbf{Q}^\mathrm{\dagger}$~\citep{calladine1978buckminster}, implying that FM and SSS form conjugate pairs. Moreover, depending on the approximation within which the system is analysed, different forms for these modes can be unveiled. For example, a Bloch wave expansion of the lattice distortions can capture all possible periodic modes and their energetic costs, with the latter contained in the lattices dispersion relation~\citep{HutchinsonR.G.2006Tspo}. There is also a certain class of affine strain producing mechanisms, commonly referred to as Guest-Hutchinson (GH) modes~\citep{guest2003determinacy}, which can be unveiled through augmenting \textbf{C} to allow for boundary deformations of the unit cell.

From \eqref{Eq:1}, Maxwell lattices can be defined in periodic boundary conditions as having an equal number of constraints and degrees of freedom, resulting in $N_\mathrm{0}=N_\mathrm{s}$. In practice, constructing a finite specimen from an infinite periodic domain will necessarily change the number of constraints, and the specimen will lose/gain FM/SSS, depending on the boundary conditions. For example, Maxwell Lattice specimens with free boundaries and a \textit{gapped} dispersion relation, will have to possess extra evanescent FM, \textit{i.e.} modes that exponentially localise on bulk discontinuities, such as free surfaces~\citep{kane2014topological}. Importantly, these edge modes can exhibit a topologically protected directional preference, also known as a topological polarization, asymmetrically changing the lattice response depending on which surface they localise. 

In this study, we will focus on the response of distorted kagome lattices. These lattice geometries can be expressed in terms of the \emph{regular kagome} (RK) lattice using four parameters $\{x_\mathrm{1}, x_\mathrm{2}, x_\mathrm{3}, z_\mathrm{s} \}$,
\begin{equation}\label{Eq:2}\textbf{r}_\mathrm{\mu} = \textbf{r}_\mathrm{\mu}^\mathrm{0}-\sqrt{3}x_\mathrm{\mu}\textbf{p}_\mathrm{\mu}+x_\mathrm{\mu-1}\textbf{a}_\mathrm{\mu+1}+\frac{z_\mathrm{s}}{\sqrt{3}}\textbf{p}_\mathrm{\mu-1},\end{equation} 
where $\textbf{r}_\mathrm{\mu}^\mathrm{0}$ are the basis vectors for the sites of the RK, $\textbf{p}_\mathrm{\mu}$ are the vectors normal to the bonds, $\textbf{a}_\mathrm{\mu}$ are the primitive lattice vectors, and $\mu\in\{1,2,3\}$~\citep{mao2018maxwell,lubensky2015phonons}. This parameterisation displaces the sites so that each of the three originally straight filaments of bonds in RK is disrupted individually by each one of the $x_\mu$, while $z_\mathrm{s}$ allows for the scaling of neighbouring triangles. A description of the lattice using the unit-cell defined by \eqref{Eq:2}, allows the topological polarization to be fully captured by a simple expression only depending on the sign of $x_\mu$~\citep{kane2014topological} when $\mathbf{a}_\mathrm{\mu}$ is fixed:
\begin{equation}\label{Eq:3} \textbf{P}_\mathrm{T} = \frac{1}{2} \sum_\mathrm{\mu=1}^\mathrm{3} \textbf{a}_\mathrm{\mu} \mathrm{sgn}(x_\mathrm{\mu}). \end{equation} As an extension of the Maxwell-Calladine index theorem, \eqref{Eq:3} encompasses information about the local count of evanescent FM and SSS concentrating on bulk discontinuities.

Our simulations are focused on finite kagome lattices with a pre-crack located along the mid-plane of the lattice, parallel to the \textit{x}-axis. Creating the pre-crack amounts to removing two constraints per unit cell, which, depending on the boundary conditions, will either introduce two FM localising on the crack edges or remove two SSS from the top/bottom boundaries of the system. Fixing $\mathbf{a}_\mathrm{\mu}$ and noting the mirror symmetry of the domain, there are only 6 non-redundant triplets of $\mathrm{sgn}(x_\mathrm{1}, x_\mathrm{2}, x_\mathrm{3})$, differentiating the relevant equivalence classes through $\textbf{P}_\mathrm{T}$. Namely (+,+,+) or (-,-,-), (-,+,+), (+,-,-), (-,+,-) and (+,-,+) which we refer to as \emph{left} (LTK) or \emph{right} (RTK) \emph{twisted kagome}, \emph{left polarised} (LP), \emph{right polarised} (RP), \emph{left-up polarised} (LUP), and \emph{right-down polarised} (RDP). We also consider the RK lattice, which represents a transition point between equivalence classes where the topological index is not defined. \rev{The simulated lattices from these equivalence classes are provided in Fig. \ref{fig:S1a}.}

\subsection{The Role of Geometry}\label{RoleOfGeometry}

Although the FM and SSS local count is topologically robust, \emph{i.e.} dictated by the system's $\mathbf{P}_\mathrm{T}$, their form can be further manipulated within this constraint. Understanding how the form of these modes depends on the exact geometrical parameters of the unit cell, beyond the established topological properties, is thus important in precisely functionalising the lattice. 

\begin{figure}
	\centering
	\includegraphics{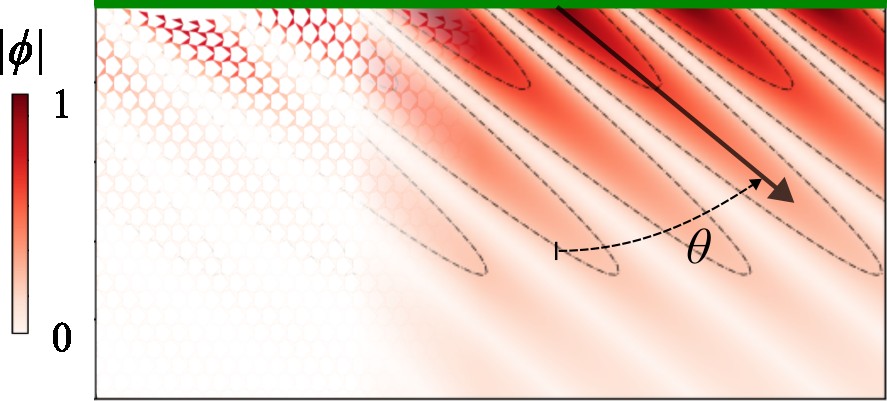} %
	\caption{Discrete and continuum surface (floppy) mode visualizations for LP1. Green solid lines represent free edges, with a distortion on the edge made up of a single wavenumber $q_\mathrm{1} = 2\pi s/N$, with a visualization of the distorted lattice fading into the continuum. The continuum plot depict the modulus $|\phi|$, of the otherwise complex FM amplitude (details of the continuum visualisation can be found in Appendix \ref{Evanescent_Mode_Appendix}).}
	\label{fig:1} 
\end{figure}

Utilising the zero-energy condition for the non-trivial FM (SSS), or equivalently observing that they fall in the nullspace of $\mathbf{C}$ ($\mathbf{Q}$), restricts their possible functional form. This constraint can be expressed in Fourier Space as
\begin{equation} \label{Eq:S5}
    E(q_\mathrm{1}, q_\mathrm{2}) = \det\mathbf{C}(q_\textrm{1},q_\textrm{2}) = \det\mathbf{Q}(q_\textrm{1},q_\textrm{2}) = 0,
\end{equation}
which amounts to fixing one wavenumber component in terms of the other, effectively reducing the degrees of freedom that are relevant for their description. This effect, known as \textit{bulk-boundary correspondence}, allows the reconstruction of the mode in the bulk given knowledge of its amplitude on a boundary (Fig. \ref{fig:1}). In essence, an evanescent mode that is described by a real wavenumber $q_\mathrm{1}$ parallel to a boundary will be associated with a wavenumber $q_\mathrm{2} \equiv q_\mathrm{2}(q_\mathrm{1})$ orthogonal to it~\citep{kane2014topological}. If the dispersion relation is gapped, the solution for the orthogonal component will be complex, $q_\mathrm{2} = k_\mathrm{2} + i\, \kappa_\mathrm{2}$, where $k_\mathrm{2}$ is the real part of the wavenumber and $\kappa_\mathrm{2}$, its imaginary part, is the inverse decay length in the $2$-direction. Therefore, a sinusoidal disturbance decays into the bulk with respect to the edge's inward facing normal at an angle given by \begin{equation}\label{Eq:4} \theta = \arctan\left(\frac{k_\mathrm{2}}{q_\mathrm{1}}\right), \end{equation} 
with clockwise rotation for a negative angle (Fig. \ref{fig:1}). As $\mathbf{C} = \mathbf{Q}^\dagger$, the $q_\mathrm{2}$ component of the SSS will be the complex conjugate of the FM resulting in $\kappa^\mathrm{SSS} = -\kappa^\mathrm{FM}$ and $\theta^\mathrm{SSS} = \theta^\mathrm{FM} = \theta$.

Although the sign of $\kappa$ is dependent on $\mathbf{P}_\mathrm{T}$, $|\kappa|$ and $\theta$ are dictated by the underlying cell geometry, allowing lattices even from the same equivalence class to have disparate forms for these modes. We will elucidate how these parameters can be leveraged between and within the equivalence classes to vary the behaviour and damage evolution of Maxwell lattices. 

Finally, the theory summarised in this section is derived assuming lattices constructed from springs that are free to rotate at their ends. Practically, it is difficult to construct a lattice with these restrictions, and in reality, non-zero bending resistance will occur. This bending resistance penalises the FM and SSS, resulting in these modes lifting from zero energy, weakening the effect of the polarization~\citep{liu2023stress} through the delocalisation effect. As per the definition of the slenderness ratio, $\mathrm{SR} = |\mathbf{a}_\mathrm{1}|/t$, where $t$ is the bond thickness and $|\mathbf{a}_\mathrm{1}|$ is the magnitude of the primitive lattice vector in the 1-direction, this quantity is identified as an additional geometrical parameter that controls the global behaviour.

\section{Numerical Method}

\begin{figure*} 
	\centering
	\includegraphics{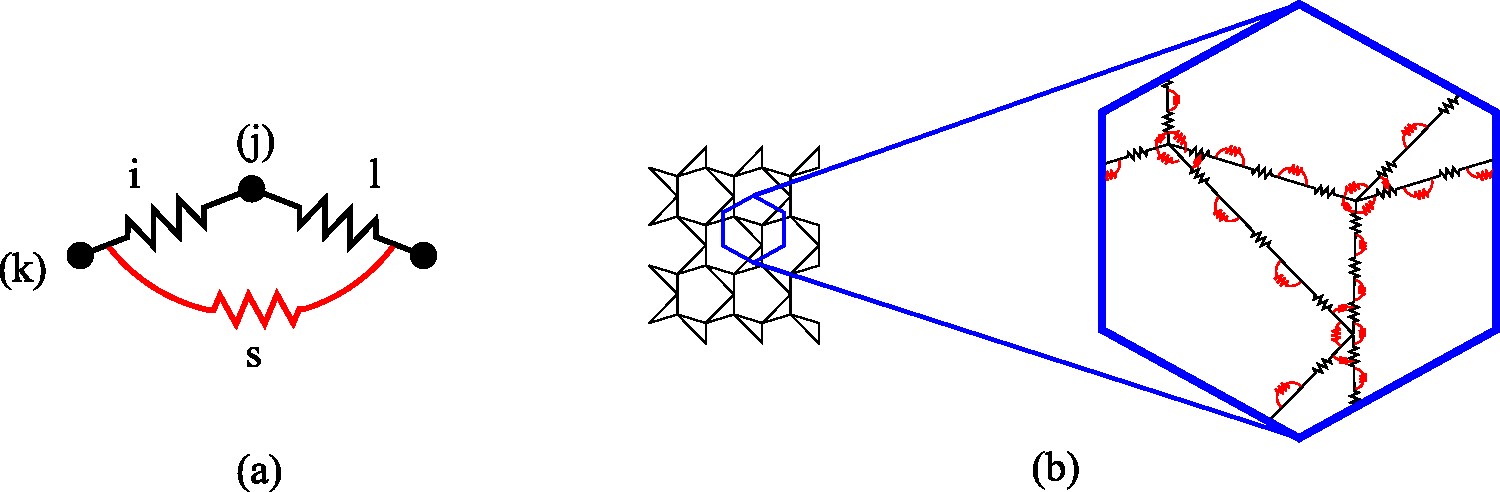} %
	\caption{(a) NN and NNN bonds. (b) Discretisation of the lattice.}
	\label{fig:2} 
\end{figure*}

In the simulations, we represent the lattice as a mass-spring network with both axial and torsional springs providing NN and NNN interactions (Fig. \ref{fig:2}a. The energy in this system is defined by 
\begin{equation}\label{Eq:6} E = \sum_{\mathrm{i=1}}^{N_\mathrm{a}} \frac{K_\mathrm{i}}{2} (l_\mathrm{i} - l_\mathrm{i}^\mathrm{0})^2 + \sum_\mathrm{s=1}^{N_\mathrm{r}} \frac{C_\mathrm{s}}{2} (\alpha_\mathrm{s}-\alpha_\mathrm{s}^\mathrm{0})^\mathrm{2}, \end{equation} 
where: $N_\mathrm{b}$ and $N_\mathrm{r}$ are the number of axial and torsional springs in the system; $K_\mathrm{i}$ and $C_\mathrm{s}$ are the axial and torsional spring stiffness (defined in Appendix \ref{ASpringStiff}); $l_\mathrm{i}$ and $l_\mathrm{i}^\mathrm{0}$ are the current length and original length of the axial spring; and, $\alpha_\mathrm{s}$ and $\alpha_\mathrm{s}^\mathrm{0}$ are the current angle and original angle between bonds $i$ and $l$. Notice that this formulation captures large deformations. $l_\mathrm{i}$ and $\alpha_\mathrm{s}$ can be written in terms of the site degrees of freedom \rev{$y_\mathrm{j}$} and, within each applied displacement step $\delta$, we minimises the force at each site ($\partial E/\partial \rev{y}_\mathrm{j}$) below a specified tolerance ($F_\mathrm{tol}$) using the FIRE 2.0 algorithm \citep{bitzek2006structural,guenole2020assessment}.

To accurately capture the bending of the bonds, we further discretise the bonds into three elements (Fig. \ref{fig:2}b). This level of discretisation was found to sufficiently capture the response of the lattice for the SR and material properties simulated. 

\begin{figure*} 
	\centering
	\includegraphics{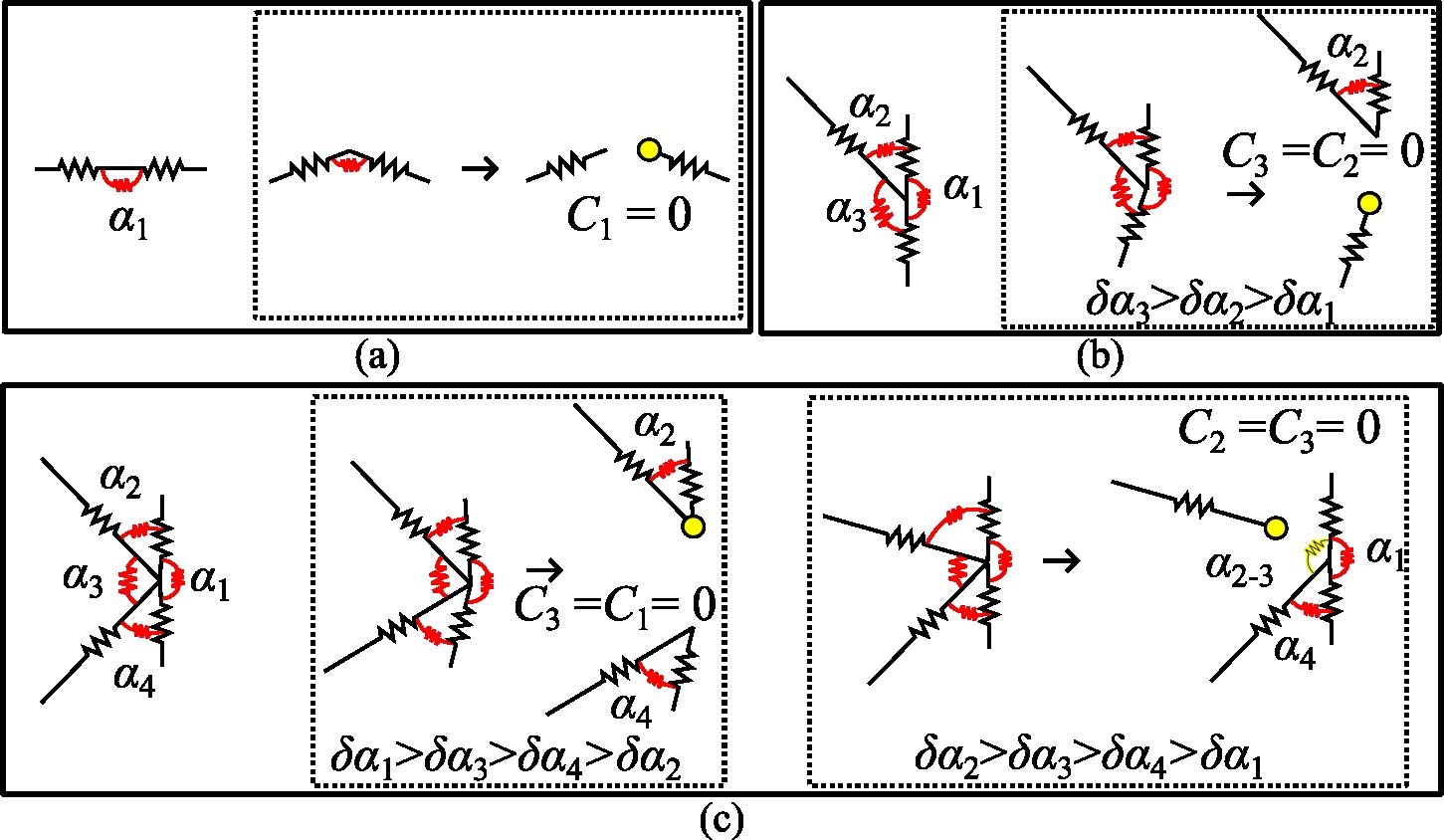} %
	\caption{\rev{Splitting of sites to simulate the fracture process. Splitting of sites with: (a) two NN bonds; (b) three NN bonds; and, (c) four NN bonds.}}
	\label{fig:3} 
\end{figure*}

An element has reached failure if 
\begin{equation}\label{Eq:7} \left|\frac{\sigma_\mathrm{A}}{\sigma_\mathrm{A}^\mathrm{u}}\right|+\left|\frac{\sigma_\mathrm{B}}{\sigma_\mathrm{B}^\mathrm{u}}\right| \geq 1, \end{equation} 
where $\sigma_\mathrm{A}$ and $\sigma_\mathrm{B}$ are the axial and bending stresses in the element (defined in Appendix \ref{AStress}), and $\sigma_\mathrm{A}^\mathrm{u}$ and $\sigma_\mathrm{B}^\mathrm{u}$ are the material failure stress values in axial and bending respectively. While only the material failure is included explicitly, buckling related failure modes are captured by this model. When a bond becomes unstable, further rotation is possible without significantly increasing the applied force \citep{el1990stress}. This far-from-threshold rotation will eventually result in the failure of an element, which is simulated by splitting the site associated with the failed element into two, creating a new edge within the lattice. When either side of the bond meets the failure criteria, we separate the bond(s) by splitting the site into two and reassigning the bonds to this new site. This is done arbitrarily when two elements are connected to a site. The associated $C_\mathrm{s}$ is also set to 0. In Fig. \ref{fig:3}a we show this process, where a yellow circle represents the new site. When three or four elements are connected at a site we choose to disconnect the elements about the two torsional springs whose angle changes the most \emph{i.e.} if two elements are rotating together about the site they will separate from the other element(s) together (Fig. \ref{fig:3}b,c). $C_\mathrm{s}$ of the two torsional springs associated with the largest rotations is then also set to 0. If, in this step, one element is separated from three elements, a new torsional spring is added to the set of three elements, connecting the adjacent elements that were previously on either side of the single separated element (Fig. \ref{fig:3}c). No torsional springs are added to the system if only two elements remain after splitting the site (Fig. \ref{fig:3}b,c). Further information about the method can be found in Appendix \ref{Method}.

To study the damage process, we simulate a rectangular domain with 24 and 16 units along the width and height respectively, where $|\textbf{a}_\mathrm{1}| = 6$ mm. At the horizontal top and bottom edge, we apply roller boundary conditions, allowing these sites to move freely parallel to the \textit{x}-axis. We also pin the left outermost site on the bottom edge to prevent any rigid body motion. Roller supports were chosen to prevent any SSS occurring at the boundary due to the additional constraints, allowing us to isolate the effect of the FM on the mechanisms of damage. The changes in the mechanisms of damage when pinned restraints are applied are subsequently studied. To apply the displacement we restrain the bottom edge from moving in the \textit{y}-direction and apply a uniform $\delta$ along the top edge. \rev{Dynamic effects are excluded to focus solely on the influence of edge states and bending rigidity.} Unless other crack lengths are specified, the pre-crack is added to the model by splitting five units from the left edge along the mid-plane of the lattice.

To investigate the effect of slenderness we simulate the lattice for six SR (2, 3, 6, 10, 20, 30). \rev{This range of SR was chosen to capture the transition from predominately stretching to bending dominated responses that fall within a realistic manufacturing range with both additive and subtractive methods}. For each SR the axial stiffness ($K_\mathrm{i}$) of the bonds is kept constant by adjusting the out-of-plane depth $b$ (\emph{i.e.} the cross-sectional area is kept constant). The change in the SR, therefore, only affects the bending rigidity ($C_\mathrm{s}$) of the bonds.

The material properties are taken from that of a relatively rigid, but brittle, material~\citep{luan2022energy}, with $E_\mathrm{\rev{m}} = 10$GPa, $\sigma_\mathrm{A}^\mathrm{u} = 65$MPa and  $\sigma_\mathrm{B}^\mathrm{u} = 126$MPa. \rev{These material properties were chosen to isolate the effect of the topological edge states and bending rigidity.} The maximum allowable residual force ($F_\mathrm{tol}$) was set to 0.0025N in the simulations following a sensitivity study.

When reporting results, the stress applied to the edge of the lattice is normalised with $\sigma_\mathrm{A}^\mathrm{u}$ (\emph{i.e.} $\sigma_\mathrm{n} = F/(A_{\mbox{\tiny edge}} \sigma_\mathrm{A}^\mathrm{u})$ where $F$ is the force applied to the sites and $A_{\mbox{\tiny edge}}$ is the projected area of the edge). The nominal strain is taken as $\epsilon_\mathrm{n} = \delta/H$, where $H$ is the initial height of the lattice.

\section{Behaviour of Topologically Distinct Lattices}\label{sec4a}

In this section we focus on how the form of FM introduced at the pre-crack edge influences the behaviour and damage evolution for mode-I opening. Initially, we focus on the results for RK, LTK, LP1, RP1 and LUP1, for which $(x_\mathrm{1},x_\mathrm{2},x_\mathrm{3})$ are (0,0,0), (0.1,0.1,0.1), (-0.1,0.1,0.1), (0.1,-0.1,-0.1) and (-0.1,0.1,-0.1) respectively. The results for the remaining equivalence class, RDP, are provided in Appendix \ref{AdditionalLattices}, along with RTK. Additionally, we include RP and LP lattices, which have chiral unit cells, \emph{i.e.} mirror-symmetric with respect to the crack plane. Results for achiral unit-cells from the same classes are also found in Appendix \ref{Achirallattices}.

\subsection{Dominant Lattice behaviour}\label{sec4}
\begin{figure*}
	\centering
	\includegraphics{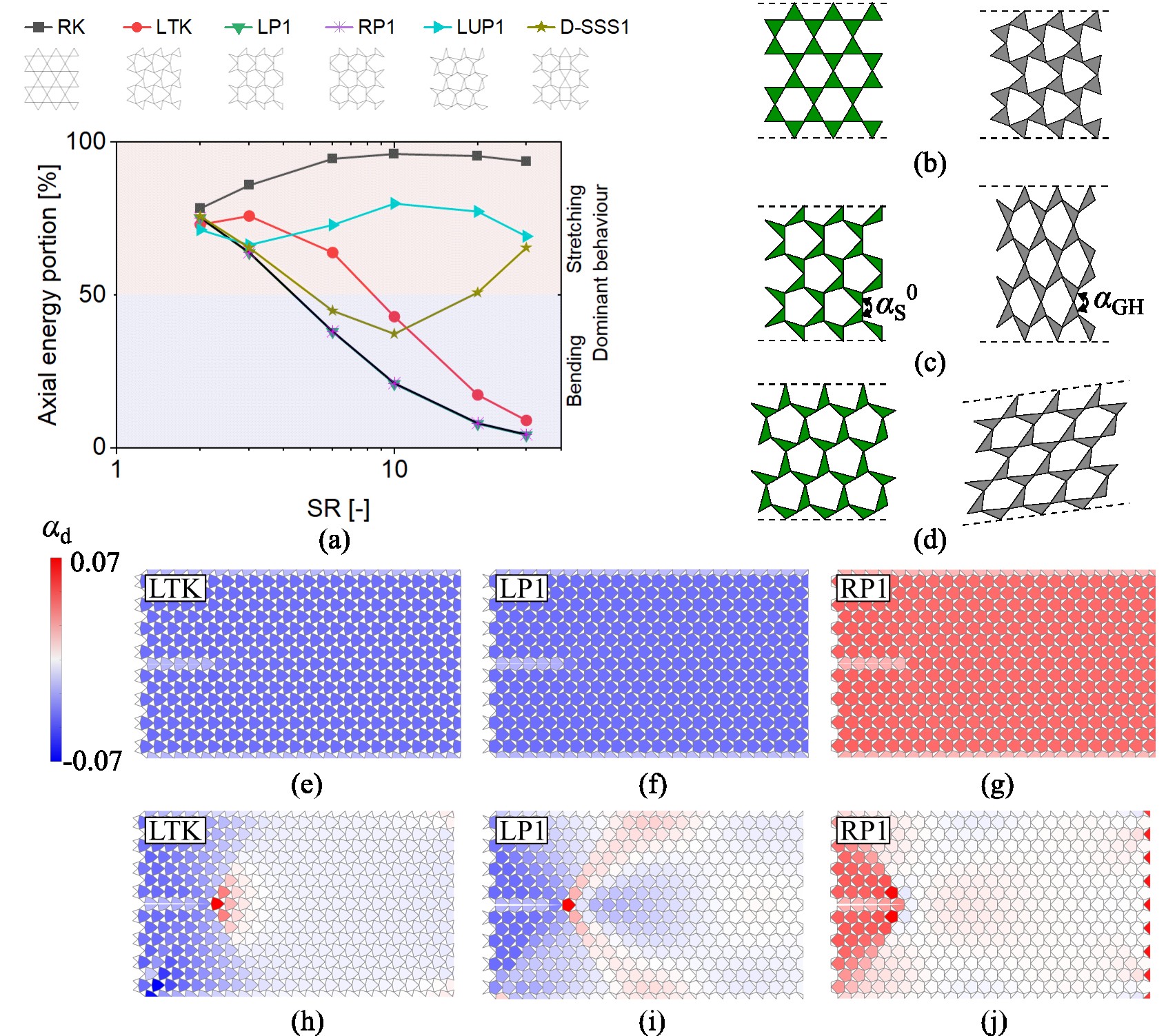} 
	\caption{\rev{Dominant behaviour of lattices. (a) Axial energy portion in the entire lattice preceding the first failure event. Visualisation of the affine strain producing mechanism (GH mode) for: (b) RK; (c) LP1; and, (d) LUP1. Comparison of the internal angle of a unit cell from the simulations ($\alpha_\mathrm{s}$) with that required by the GH mode ($\alpha_\mathrm{GH}$) to displace the boundary the equivalent distance (\emph{i.e.} $\alpha_\mathrm{d} = \alpha_\mathrm{s}-\alpha_\mathrm{GH}$) for: (e, h) LTK, (f, i) LP1; and, (g, j) RP1 at SR = 3 and 30 respectively. The difference in angle is visualised through the colour directly to the right of each unit cell. The simulations for these lattices are included in movie S1.}}
	\label{fig4}
\end{figure*}

Subjecting all lattices to mode-I opening, \emph{i.e.} displacing the horizontal boundaries normal to the crack plane, gives rise to different dominant contributions for the stresses in the bulk, depending on the geometry of the unit-cell. With reference to Fig.~\ref{fig4}a, all lattices are found to be stretching-dominated at low SR values (stocky beam elements), but their behaviour diverges significantly at larger SR (slender beam elements), as the bending resistance of the unit-cell decreases. 

In practice, bonds are constrained from freely rotating around each site due to the local bending resistance associated with finite thickness joints. However, contrasting the idealised GH modes (\emph{i.e.}, affine strain-producing mechanisms) with the lattice deformation in the simulator, offers insights into the mechanics while traversing SR-space. \rev{Geometries that exhibit a GH mode will, in practice, become bending-dominated at high SR values when the applied boundary displacements closely correspond to those demanded by the GH mode (\emph{e.g.}, Fig. \ref{fig4}c).} The LTK, LP, and RP lattices, all belong in this category. \rev{Conversely, lattices without a GH mode, such as RK, or whose GH mode yields boundary displacements incompatible with the simulations, such as LUP (Fig.~\ref{fig4}d), remain stretching-dominated across the entire SR range.} Here, there is a caveat for the RK lattice, whose uniform FM depicted in Fig.~\ref{fig4}b is not classified as a GH mode, as it does not produce any affine strain to linear order in the displacements. This mode is rather an infinitesimal increment of the GH mode \citep{HutchinsonR.G.2006Tspo}, which can only map to boundary displacements in compression through an instability, leading to stretching-dominated behaviour in tension across the SR range.

\begin{figure*}
	\centering
	\includegraphics[width=1.0\textwidth]{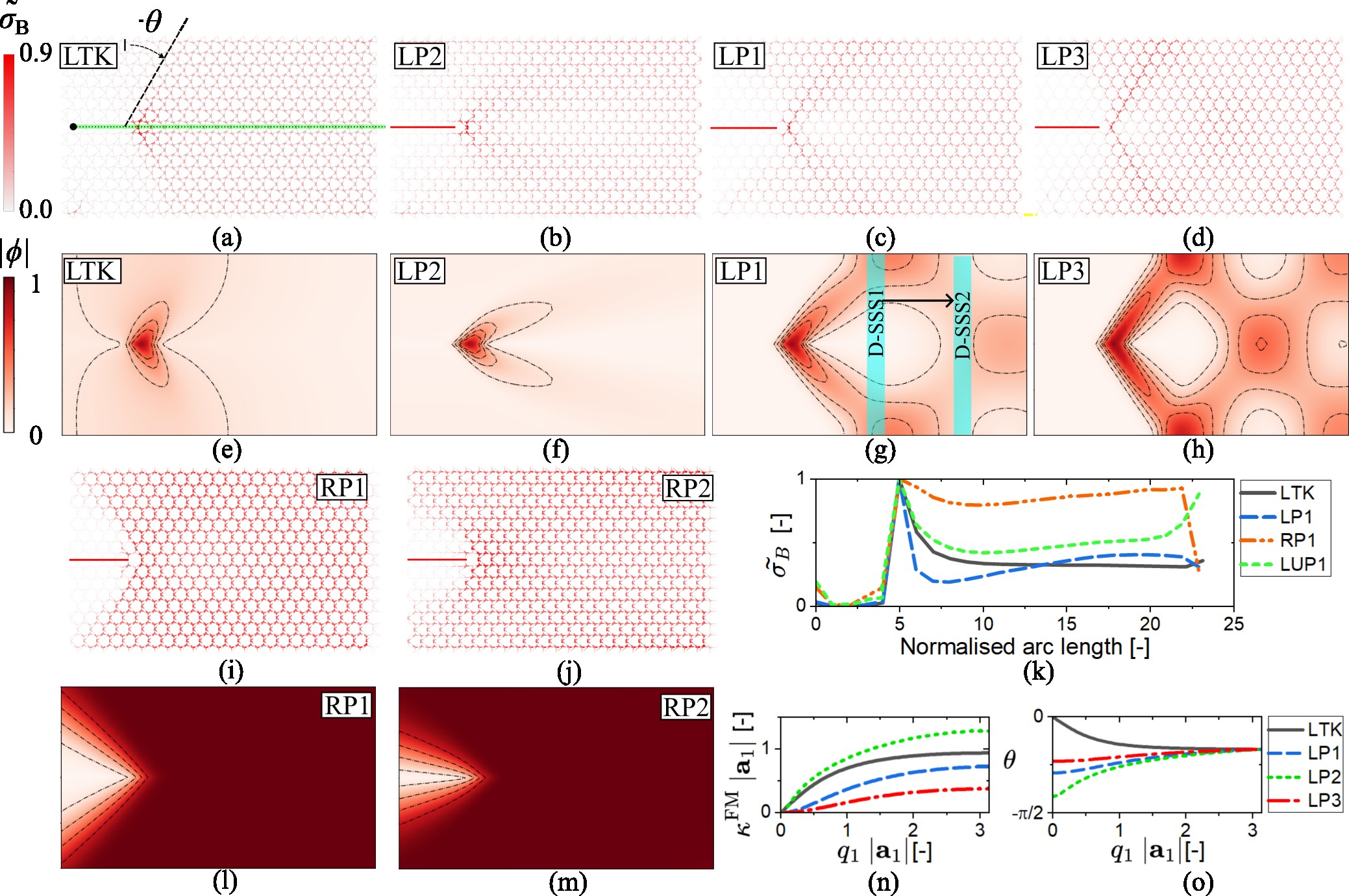} 
	\caption{\rev{Mapping of bending stresses through the bulk-boundary correspondence. Maximum normalised bending stress in the elements cross section $\sigma_\mathrm{\textbf{Bend}} = |\sigma_\mathrm{B}/\sigma_\mathrm{B}^\mathrm{u}|$ and visualisation of FM through the magnitude of the total modulation ($|\phi|$) of the wave spectrum for: (a, e) RK (0.1,0.1,0.1); (b, f) LP2 (-0.02,0.18,0.18); (c, g) LP1 (-0.1,0.1,0.1); (d, h) LP3 (-0.16,0.04,0.04); (i, l) RP1 (0.1,-0.1,-0.1); and, (j, m) RP2 (0.02,-0.18,-0.18). (k) Normalised bending stress along the arc length. The bending stress is normalised with the peak bending stress along this length. The arc length is illustrated in (a). (n) Decay spectrum of zero modes on a horizontal surface; and, (o) associated wave angles. All figures associated with simulations are for SR = 30. The simulations for these lattices are included in movie S1-2.}}
	\label{fig5} 
\end{figure*}

To explore whether non-idealised lattices deform along the GH mode, further visual comparisons are presented for low (Fig. \ref{fig4}e-g) and high (Fig. \ref{fig4}h-j) SR. These figures highlight the difference between the simulated local angle, $\alpha_\mathrm{s}$, and the expected GH angle, $\alpha_\mathrm{GH}$. Despite the majority of the bulk following the GH mode for the bending-dominated lattices at $\text{SR} = 30$, they also posses regions that diverge from this behaviour. Bending stresses mirror this trend (Fig.~\ref{fig5}a,c,\rev{i}), with higher stress values appearing in the same regions as the maps for $\alpha_\mathrm{d}$ in Fig.~\ref{fig4}. Below, we argue that the form of the regions that diverge from the GH mode can be connected to the edge actuation through the bulk-boundary correspondence. Thus, the phenomena observed can be interpreted in light of the compliance added to the system by the pre-crack.

The bending stresses along the mid-plane (Fig. \ref{fig5}\rev{k}) reveal the FM actuation both at the crack edge and within the lattice. For LTK and LP1, the actuation is mainly concentrated in a sharp peak at the crack tip, while for RP1, the peak remains close to the uniform value observed in the bulk. All lattices are relaxed along the rest of the crack edge, to the left of the crack tip. By modelling the FM as a modulation of the uniform zero-energy GH mode~\citep{rocklin2017transformable}, without restricting the description to linear order, we can leverage bulk-boundary correspondence, prescribing an analogue of the observed dominant actuation pattern along the mid-plane, to visualize the resulting FM amplitude, $|\phi|$, in the bulk. In particular, we prescribe a \emph{Gaussian} spike of unit height and width, centred at the crack-tip, for the LP/LTK lattices (Fig.~\ref{fig5}e-h), and a \emph{sigmoid} function centred at the crack-tip for RP (Fig.~\ref{fig5}\rev{l,m}). Different prescriptions can be superimposed, so one could easily add the Gaussian and sigmoid solutions together with different relative amplitudes, to reach a more accurate visualisations of the FM in the bulk for each lattice. However, the goal here is to capture the most dominant components of the disturbance; the method should be seen as a simple heuristic (further details of the visualisation can be found in Appendix \ref{Evanescent_Mode_Appendix}).

Three additional lattices within the LP/RP ($\mp$,$\pm$,$\pm$) equivalence classes are simulated to further illustrate how the form of the FM influence the stress profile within the lattice. Even with the simplified aforementioned prescriptions, the FM fields in the bulk (Fig.~\ref{fig5}e-h,\rev{l,m}) map qualitatively well onto the major components of the bending stresses in the lattice (Fig.~\ref{fig5}a-d,\rev{i,j}). To better understand the form of the FM in the bulk, Fig.~\ref{fig5}\rev{n,o}, can offer valuable insights. Firstly, the penetration into the bulk of the localised disturbance can be connected to the decay rate $\kappa^\mathrm{FM}(q_\mathrm{1})$ spectrum. At the homogenisation limit (\emph{i.e} small $q_\mathrm{1}$), non-polarised lattices (\emph{e.g} LTK) are expected to decay faster because they have $\kappa^\mathrm{FM}(q_\mathrm{1}) \sim \mathcal{O}(q_\mathrm{1})$, compared to $\kappa^\mathrm{FM}(q_\mathrm{1}) \sim \mathcal{O}(q^2_\mathrm{1})$ for polarised lattices \citep{kane2014topological}. Here, we observe that, in general, the magnitude of $\kappa^\mathrm{FM}(q_\mathrm{1})$ across the whole range is more indicative, which makes sense for a localised disturbance, as it samples a wide spectrum in $q_\mathrm{1}$-space. Subsequently, LTK and LP2 decay quickly into the bulk, with the deformation away from the crack-tip reverting back to that associated with the idealised GH mode. Contrarily, the spectra of LP1 and LP3 exhibit lower decay rates, with the disturbances influencing the behaviour deeper into the bulk. 

As the large $q_\mathrm{1}$ components of the mode decay faster, the direction of propagation for the disturbance far from the crack tip is dictated by the small $q_\mathrm{1}$ limit of $\theta(q_\mathrm{1})$. This is clearly discernible by comparing, \emph{e.g.}, LP2 and LP3, where the propagation for the former happens at an angle much closer to the horizontal. Additionally, the form of the extra modes for RP, which propagate towards the left edge (positive $\theta$), do not offer an alternative deformation mechanism for the units within the bulk. The only mode available for these is associated with the uniform GH mode, leading to no significant stress localization at the crack tip. However, notable reduction in bending stress is observed along the right edge of RP1-2 (Fig. \ref{fig5}\rev{i,j}), due to that edge's FM which locally increase the compliance.

The dominant mode actuation results in the strength difference observed between RP1 and LP1 at SR = 30 (Fig.~\ref{fig8}b). The deformation for RP1 is more uniform, closely following the GH mode within the bulk. In contrast, for LP1, the deformations localise along the FM, resulting in higher stresses for the equivalent applied displacements and, therefore, lower peak stress. As the SR decreases, the deformation of the lattice is less influenced by these FM (due to the associated increase in the bending resistance), resulting in a reduction in this difference. 

Finally, the stress within the bulk of stretching-dominated lattices (\emph{i.e.}, LUP1) is not significantly affected by the FM introduced at the crack edge. However, at the crack-tip the FM are actuated in a localised manner (Fig.~\ref{fig5}\rev{k}) and these greatly influence the failure, which will be discussed in the following section. 

\subsection{Dominant Lattice Failure}

\begin{figure} 
	\centering
	\includegraphics[scale=0.90]{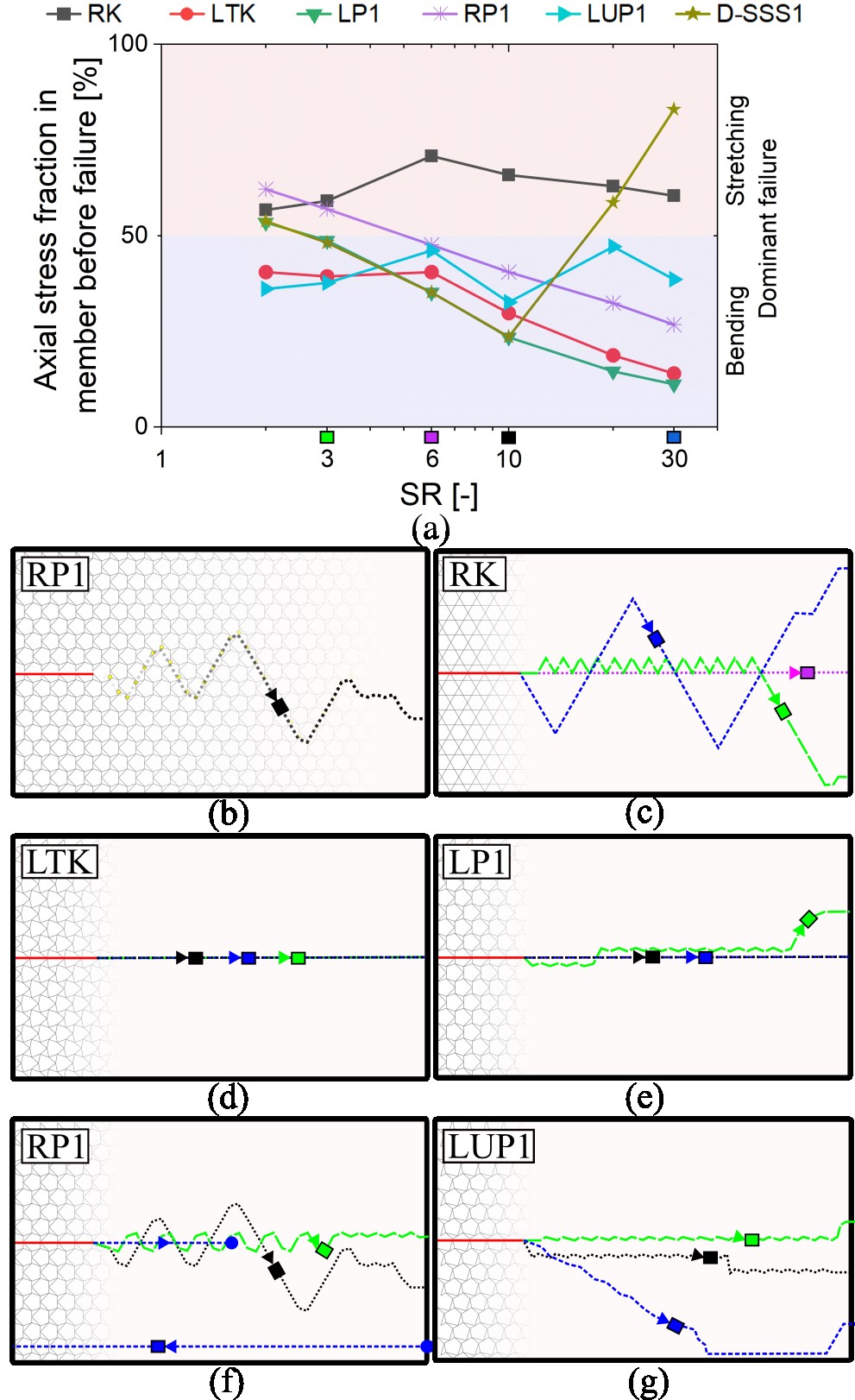} 
	\caption{Dominant failure of lattices. (a) Axial stress fraction in member that fractures preceding the first failure event. (b) The damage process is visualised with a line traversing the domain, connecting separated nodes (fracture locations) within the lattice, indicated by yellow markers. For clarity we do not show the microstructure throughout the domain. The damage process for SR = 3, 6, 10, 30 are plotted with green, purple, black, and blue lines and associated rectangular markers respectively (refer to (a)). Arrows indicate the direction of damage and circular markers indicate a discontinuity in the damage path. Damage paths for: (c) RK; (d) LTK; (e) LP1; (f) RP1; and, (g) LUP1. The simulations for these lattices are included in movie S1.}
	\label{fig6} 
\end{figure}

Fig.~\ref{fig6}a provides the axial stress fraction in the element that fails in the step immediately before fracture. The damage paths, provided in Fig.~\ref{fig6}b-g, each showing results for a range of SR, appear to be influenced by the proportion of axial and bending stress within the element. When bending or stretching dominates the fracture, the crack appears to propagate horizontally parallel to the pre-crack edge for symmetric about the mid-plane lattices, \emph{i.e.} RK at SR = 6, LTK at SR = 2-30, LP1 at SR = 6-30, and RP1 at SR = 20-30. As the stress contributions become similar, the damage path generally deviates from the pre-crack plane as it traverses the lattice. 

\begin{figure*} 
	\centering
	\includegraphics{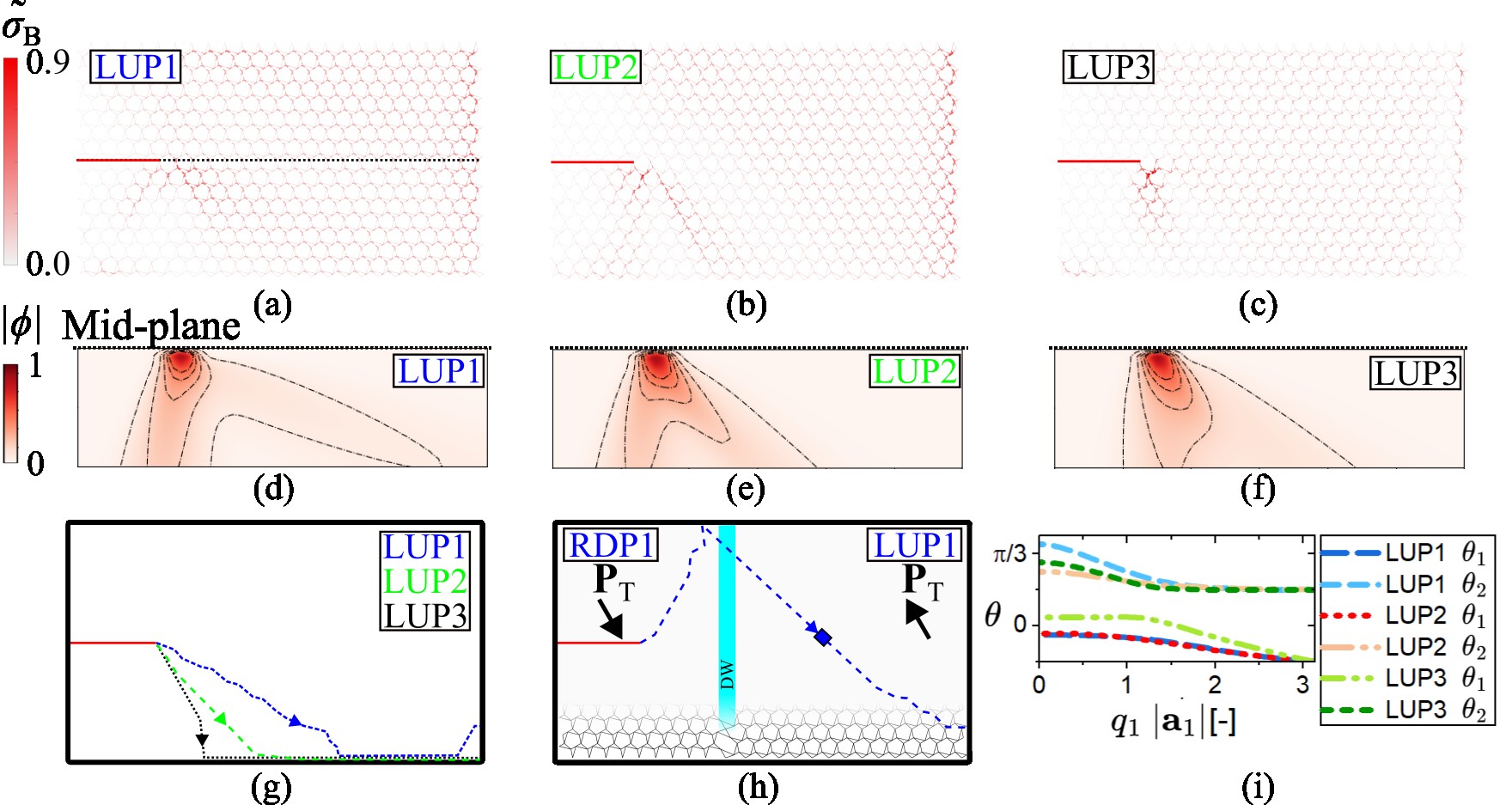} 
	\caption{\rev{Damage path control using FM. Maximum normalised bending stress in the elements cross section $\sigma_\mathrm{\textbf{Bend}} = |\sigma_\mathrm{B}/\sigma_\mathrm{B}^\mathrm{u}|$ and visualisation of FM through the magnitude of the total modulation ($|\phi|$) of the wave spectrum for: (a, d) LUP1 (-0.1,0.1,-0.1); (b, e) LUP2 (-0.18,0.04,-0.1); and, (c, f) LUP3 (-0.16,0.04,-0.16). (g) Damage path for LUP1, LUP2, LUP3 at SR = 30. (h) Damage path for lattice with a domain wall separating RDP1 and LUP1. (i) $\theta$ spectrum for LUP1, LUP2 and LUP3. All figures associated with simulations are for SR = 30. The simulations for these lattices are included in movie S3.}}
	\label{fig7} 
\end{figure*}

While the LUP1 global behaviour is stretching-dominated, the fracture of its elements is strongly influenced by the bending (Fig.~\ref{fig6}a) associated with the FM localising at the pre-crack edge at large SR (Fig.~\ref{fig7}a,d). As such, strong alignment of the damage path with the long-wavelength limit of $\theta_\mathrm{2}$ (Fig.~\ref{fig7}i) is observed. Simulating other lattices from the equivalence class (-,+,-) at SR = 30, it indicates this is a robust phenomenon: for LUP2 (-0.18,0.04,-0.1) and LUP3 (-0.16,0.04,-0.16) the damage path (Fig.~\ref{fig7}g) follows the long-wavelength limit of $\theta_2$ and $\theta_1$ respectively (Fig.~\ref{fig7}d-f,i). This results in the damage evolution following a shallower angle than LUP1. Close to the long-wavelength limit, $\theta_1$ propagates further into the bulk for LUP1-3. While the lattice appears to favour propagating the crack in the direction with the smallest $\kappa^\mathrm{FM}$ (\emph{i.e.} LUP3), if $\theta$ is behind the crack tip, it instead favours the modes that propagate forward, ahead of the crack tip (\emph{i.e.} LUP1 and 2). This phenomenon is also observed for the RDP lattices, provided in Appendix \ref{AdditionalLattices}. For lattices in the ($\pm$,$\mp$,$\pm$) equivalence classes, at large SR, there is a strong alignment of the damage path with the FM localised at the crack tip, providing a control parameter to steer the localised fracture of its underlying microstructure. As the SR reduces, the crack tends towards a horizontal path (Fig.~\ref{fig6}g), with these FM having a reduced influence on the damage.

\section{Effect of Discontinuities}

In the previous section, we showed that fundamental to the kagome lattices response is the form of the FM, and we demonstrated how these can be manipulated between and within the same equivalence class to influence the stress profile and damage evolution. In this section, we now examine how the behaviour of the lattice is affected by two types of discontinuities: (1) a topological domain wall; and, (2) additional constraints along the edge of the domain. 

\subsection{Topological domain walls}

\begin{figure*} 
	\centering
	\includegraphics{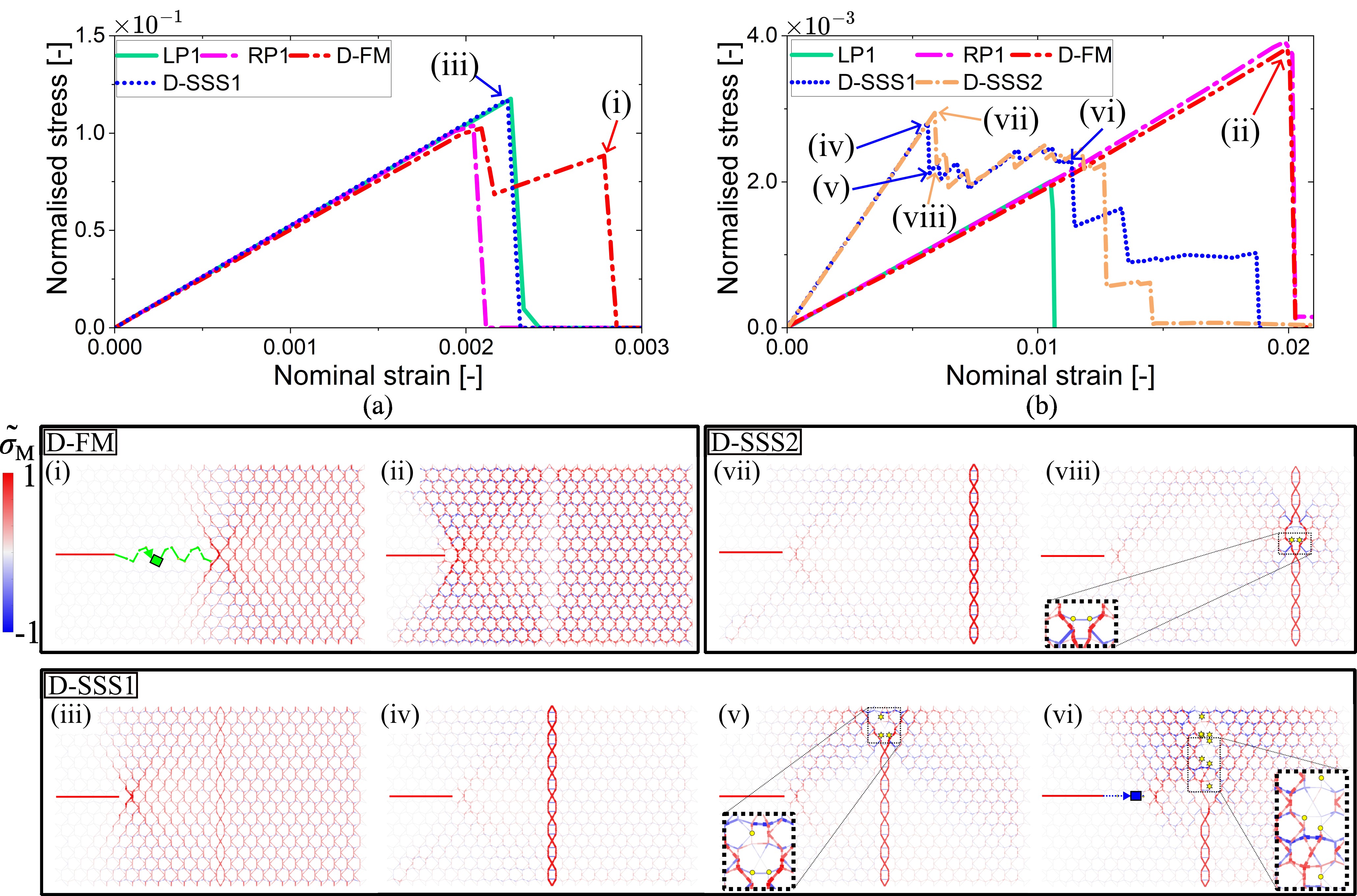} %
	\caption{Effect of domain walls. Normalised applied stress vs nominal strain for: (a) SR = 3; and, (b) SR = 30. Maximum normalised stress in the elements cross section for D-FM, D-SSS1 and D-SSS2 at specified nominal strains for: (i,iii) SR = 3; and, (ii,iv-viii) SR = 30. Where $\sigma_\mathrm{M} = \sigma_\mathrm{A}/\sigma_\mathrm{A}^\mathrm{u} \pm |\sigma_\mathrm{B}/\sigma_\mathrm{B}^\mathrm{u}|$. In (v), (vi), and (viii) yellow stars indicated nodes that are added into the system due to the separation of elements (element fracture) along the domain wall. For clarity, an insert is also included where circular yellow markers indicate these nodes. The simulations for these lattices are included in movie S4.}
	\label{fig8} 
\end{figure*}

Combining LP1 and RP1 lattices results in a localised \emph{domain wall} of FM (D-FM) or SSS (D-SSS1). These domain walls are placed between the 13$^\mathrm{th}$ and 14$^\mathrm{th}$ unit cell columns. The normalised peak stress for D-FM remains similar to RP1 throughout the SR range (Fig.~\ref{fig8}a,b), indicating the FM domain wall does not significantly influence the stress profile around the crack when placed in this position. However, the initial stiffness of D-FM is marginally smaller than that of RP1, which is due to the FM localising at the domain wall relieving stresses. This is clearly observed in Fig.~\ref{fig8}ii.  

Pairing lattices from different equivalence classes provides a mechanism to arrest the crack at the domain wall. At SR = 3, LP1 has a higher peak stress than RP1 (Fig.~\ref{fig8}\rev{a}). Once initiated, the crack traverses D-FM until it meets the domain wall (Fig.~\ref{fig8}i). The load then increases before the damage continues, resulting in the stiffening behaviour observed in Fig.~\ref{fig8}a. In contrast, at SR = 30, RP1 has a higher peak stress, resulting in the damage propagating unstably through D-FM with a sudden reduction in stress (Fig.~\ref{fig8}\rev{b}). This stiffening is also observed for D-SSS1 and D-SSS2 once the damage reaches the SSS domain wall (refer to movie S4).

At low SR, the SSS at the domain does not develop due to the delocalisation associated with the increased bending rigidity. As a result, the response of D-SSS1 at SR = 3 is almost identical to LP1 (Fig.~\ref{fig4}a, Fig.~\ref{fig6}a and Fig.~\ref{fig8}a), with stress concentrating at the crack tip rather than at the domain wall (Fig.~\ref{fig8}iii). At SR = 30, the response drastically changes, where D-SSS1 has a significantly larger initial stiffness (Fig.~\ref{fig8}b) and high stresses localise at the domain wall (Fig.~\ref{fig8}iv). In this initial stage, the lattice behaviour is stretching-dominated (Fig.~\ref{fig4}a), and instead of failing at the crack tip, the failure initially localises at the domain wall---several fracture events occur before damage propagates from the pre-crack. This results in the stiffening observed in Fig.~\ref{fig8}b. At SR = 20, a similar response is observed, albeit the stress delocalises further from the domain wall. Reducing the SR further to 10, results in the fracture initiating from the crack tip, with the dominant failure converging to LP1 (Fig.~\ref{fig6}a). To understand the origin of this dependence in SR, it is relevant to recognise that deformation is required in the bulk to develop the SSS at the domain wall. When joint rotations away from the domain wall accommodate this deformation, due to their low bending rigidity at large SR, the stresses at the crack tip are not too large, and they rather develop along the domain wall. Whilst the damage path for D-SSS1 and LP1 initiates at the crack tip for SR = 10, the dominant behaviour for D-SSS1 has a larger stretch portion than LP1 (Fig.~\ref{fig4}a). This difference reduces with the SR and is also related to the degree to which the stress in the domain wall develops. 

The locations of failure along the SSS domain wall map onto the locations where FM emanating from the crack tip intersect it. Fig.~\ref{fig5}g highlights the regions where the SSS domain wall is placed with respect to the FM, and indeed, the failure initially localises where the FM cross the domain wall for both D-SSS1 (Fig.~\ref{fig8}v) and D-SSS2 (Fig.~\ref{fig8}viii). The location of the highest stress in the domain wall can therefore indicate the location of any pre-crack/defect within the lattice. By placing sensors along the domain wall, and understanding the form of the FM propagating from a crack, the stresses along the domain wall may be used to triangulate these defects. Placing the SSS in the region with less FM variation also increases the symmetry in the stress developing along the domain wall, prolonging failure and increasing the peak stress for D-SSS2 when compared to D-SSS1 (Fig.~\ref{fig8}b). 

Controlling the damage path is also possible through domain walls. For instance, combining LUP1 and RDP1 results in the damage propagating diagonally upwards and downwards before and after reaching the domain wall (Fig.~\ref{fig7}h). This can be leveraged to guide the crack as it propagates through the lattice, providing a framework to increase the total fracture energy and/or steering the damage away from valuable components.

\subsection{Additional constraints}
Applying pinned boundary conditions adds constraints to the system, which will either remove FM and/or introduce SSS localised at the boundary. The addition of these SSS can alter the stress profile within the lattice, and in this section, we will show how the form of these influence the mechanisms of damage.

\begin{figure*}
	\centering
	\includegraphics{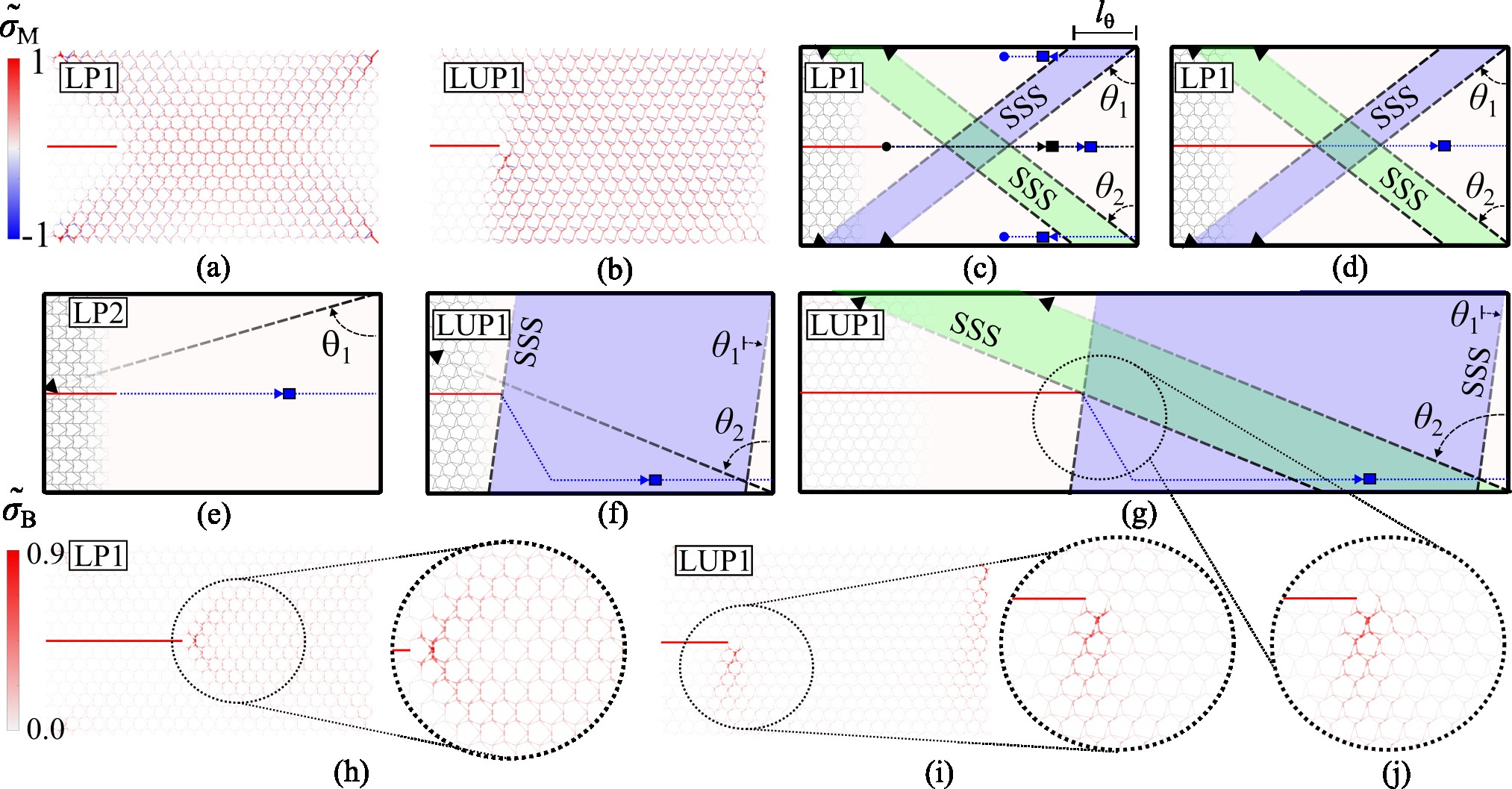} 
	\caption{Effect of pinned boundary conditions. Maximum normalised stress in the elements cross section for (a) LP1 and (b) LUP1 at SR = 30. Damage path with pinned boundary for: (c) LP1 at SR = 10 and 30; (d) LP1 at SR = 30 with 10 unit pre-crack length; (e) LP2 at SR =30; (f) LUP1 at SR = 30; and, (g) LUP1 at SR = 30 with 50 units along the length. In (c-g) we also highlight the zones where the SSS at the long-wavelength limit may develop with arrows either side of these zones. If the SSS can't develop, we only plot the direction of the SSS with a single arrow. Maximum normalised bending stress in the elements cross section at SR = 30 for: (h) LP1 with 10 unit pre-crack length; (i) LUP1; and, (j) LUP1 with 50 units along the length. The simulations for these lattices are included in movie S5.}
	\label{fig9} 
\end{figure*}

A highly localised actuation for the pre-crack FM, producing a spike in the deformation field, resulted in a wide spread for the spectrum of the mode in Fourier space. In contrast, the SSS along the boundary may spread over a larger length of the boundary $l_\mathrm{\theta}$ (Fig.~\ref{fig9}c), resulting in a spectrum focused on modes close to the long-wavelength limit, which possess a low decay rate \emph{i.e.} $\kappa^\mathrm{SSS}\rightarrow 0$ as $q_\mathrm{1}\rightarrow 0$. Thus, at the homogenisation limit, the length of the boundary on which the respective SSS can develop, $l_\mathrm{\theta}$, can be given by
\begin{equation}\label{Eq:5} l_\mathrm{\theta} = L-H\tan{\left[\theta(q_\mathrm{1}=0)\right]},\end{equation} 
where $H$ and $L$ are the lattice domain height and length respectively. If this value is negative, there can be no SSS localised on that boundary. In Fig.~\ref{fig9}c,d,f,g we highlight SSS zones at the long-wavelength limit that may develop within the respective lattices.

The SSS for LP1, decaying diagonally left from the top and bottom edges into the bulk, aligns significantly with the applied boundary conditions. As a result, the axial energy portion at SR = 30 jumps from 4$\%$ with roller supports to 59$\%$ when pinned boundary conditions are applied. A significant correlation between the stress profile found in the simulations and the long-wavelength limit SSS zones is observed when comparing Fig.~\ref{fig9}a and c. Significant stretching occurs at the right of the top and bottom edges, with lower stressed zones matching the regions where the SSS zones do not occur. At the right edge, SSS following the short-wavelength limit also appear; however, as these have the largest $\kappa_i^\mathrm{SSS}$, they decay before reaching the opposite edge. 

The lattice size should not significantly affect its dominant behaviour when the SSS at the long-wavelength limit dominate the mechanical response. This occurs when $l_\mathrm{\theta}$ is above a critical length that localises the SSS towards this limit, as these modes do not decay. In Appendix \ref{PinnedSizeEffects} we simulate a series of LP1 lattices with increasing domain size but identical aspect ratio. Keeping the aspect ratio constant ensures the SSS zones relative size remains constant. Plotting the axial energy portion for these lattices shows that the dominant behaviour does not vary significantly, and therefore, for this aspect ratio the SSS associated with the long-wavelength limit dominates the response. 

When SSS develop within the lattice, the location of damage initiation is greatly influenced by the pre-crack size, particularly at large SR. For LP1, the pre-crack spans 5 units in length and does not disrupt the SSS zones within the lattice (Fig.~\ref{fig9}c), resulting in low stresses developing here (Fig.~\ref{fig9}a). Instead, the damage initiates where the SSS localise, with the damage continuing 10 units inwards along these edges before propagating horizontally from the pre-crack edge (Fig.~\ref{fig9}c). When the crack is extended to disrupt the SSS, the deformations become localised at the crack tip, and damage initiates here rather than at the top and bottom edges (Fig.~\ref{fig9}d). In this case, the bending stress about the crack tip is similar to that observed with roller constraints (Fig.~\ref{fig5}c), although they significantly reduce as they approach the pinned boundary (Fig.~\ref{fig9}h). The additional constraints lift the energy of FM that have not decayed sufficiently on reaching the boundary. However, if the domain is large enough, the stress profile around the crack is expected to follow the actuated FM form. To prevent the debonding type damage, when the crack length does not disrupt the SSS zones, we can adjust the geometry so that the SSS at the edges do not develop fully. This can be achieved by reducing the SR, where for SR = 10, no fracture events occur at the edges where the boundary conditions are applied (Fig.~\ref{fig9}c).

A similar damage evolution is observed in other confined non-topologically polarised lattices when $z > 2d$ \citep{hedvard2024toughening}. In these lattices, an incomplete debonding type failure occurs, leaving bridging elements along the interface. Subsequently, these elements fracture, completely debonding the interface. SSS are expected to form within the lattice as it is confined and $z > 2d$. Therefore, this initial failure process may occur due to the removal of these SSS. Once removed, the remaining bridging elements become bending-dominated (\emph{i.e.} the local $z \leq 2d$), leading to the observed toughening behaviour. For lattices with $z < 2d$ this bridging is not as readily observed \citep{hedvard2024toughening}.  

When the size of the lattice prevents the SSS from developing, the dominant behaviour and failure do not vary greatly between roller or pinned boundary conditions. This is the case for LP2, where the damage path is identical for both boundary conditions (Fig.~\ref{fig9}e) and the axial energy only rises from $3\%$ to $11\%$. LP2 would require a minimum length of 49 units for the SSS at the long-wavelength limit to develop. Increasing the lattice length to 50 units results in these SSS developing, and in Appendix \ref{LP2Larger} we show that the fracture initiates at the top and bottom edges. The axial energy portion further increases to $22\%$, highlighting that the original length of LP2 was not large enough for these SSS to develop.

For LUP1 the axial energy portion goes from approximately 70\% to 95\% when the boundary conditions are switched to pinned for SR = 30. The SSS long-wavelength limit zones shown in Fig.~\ref{fig9}f strongly correlate to the stress that develops within the lattice (Fig.~\ref{fig9}b). The SSS corresponding to $\theta_1$ and $\theta_2$ require 2 and 34 units along the length to develop for the domain size simulated. The SSS associated with $\theta_2$, therefore, does not occur. The change in the stress profile with respect to the form of these modes is provided in Appenix \ref{LUPPinned}  for LUP1-3.

When the SSS at the long-wavelength limit dominates the response within the bulk, the FM activated due to the pre-crack also tend towards the associated complex conjugate of these modes in LUP1. The damage, therefore, propagates at a much shallower angle for LUP1 (Fig.~\ref{fig9}f) when compared to roller constraints (Fig.~\ref{fig7}g). Rather than stress around the crack tip following the FM at the long-wavelength limit of $\theta_2$, they instead follow $\theta_1$ (Fig.~\ref{fig9}i), as this is the FM associated with the dominant SSS. Increasing the length of the lattice so that SSS associated with $\theta_2$ can develop; and, by increasing the crack length so it intercepts this SSS, we show that bending stresses in the direction of $\theta_2$ also visually increase around the crack tip (Fig.~\ref{fig9}j). However, as the SSS associated with $\theta_1$ overlap a larger degree with the boundary conditions applied, the FM associated with $\theta_1$ continue to dominate, and the damage propagates identically to that observed for the smaller lattice domain length (Fig.~\ref{fig9}g).

\section{Conclusions}

Designing the mechanisms of damage in metamaterials has predominately leaned on the inclusion of heterogeneities to overcome the plethora of influencing parameters and competing length scales. In this work, we instead show that homogeneous Maxwell lattices have topological and geometry-dependent parameters that dominate the locations of stress concentrations, particularly at large SR. \rev{These parameters arise from the form of evanescent FM and SSS, which themselves derive from bulk geometrical properties, hence providing a design framework that does not rely on heterogeneities, and is agnostic with respect to the location of crack initiation.} \rev{Furthermore, by finely controlling the form of these modes, \emph{i.e.} their decay and angle of propagation, we demonstrate how our framework can be used to design the response of lattices without necessarily changing their topological properties.}
Using a beam element simulator, we numerically probed the effect of these modes over a range of slenderness for a set of distorted kagome lattices that span the possible equivalence classes. 

At large SR, the dominant behaviour of these lattices was found to be dependent on the boundary conditions applied. If the boundary conditions map onto the idealised uniform zero-energy macroscopic strain producing mechanism \emph{i.e.} the GH mode, the lattice bonds will preferentially bend. Conversely, when they do not, the bonds will preferentially stretch. At low SR, all lattices become stretching-dominated as the energetic cost to bend the bonds becomes unfavourable. 

When the lattice is bending-dominated, we show that the bending stress distribution within the lattice can be predicted through the bulk-boundary correspondence, which determines the amplitude of a disturbance in the bulk given a prescription on a boundary. For stretching-dominated lattices at \rev{large} SR, while the stress profile within the bulk of the lattice is not governed by the FM that localise at the discontinuity, the bending stresses around the crack tip are. Consequently, when the fracture in these lattices is governed by bending, the damage strongly aligns with these modes. At low SR, the effect of these modes was found to diminish due to the delocalisation of the stress associated with increasing bending rigidity. 

Moving away from a homogeneous domain, we combine lattices from disparate equivalence classes to form domain walls and show how differences in strength between the lattices can be leveraged to arrest the damage propagation at the domain wall across the SR range. Additionally, by combining these distinct lattices to form SSS domain walls, the fracture can initially be localised at the domain wall rather than at the pre-crack for large SR. Notably, we show the location where the fracture initiates along the domain wall coincides with locations where the FM that propagate from the pre-crack edge intersect it. Furthermore, by leveraging the mapping between the form of the FM and the damage path at large SR, we show how lattices can be combined to steer the damage through the domain.

Finally we probe the effect of pinned boundary conditions. In this scenario we show the FM conjugate partner, the SSS, dominates the response at large SR. Provided the domain aspect ratio allows the SSS to develop, we show that the SSS at the long-wavelength limit accurately predicts the stress profile within the lattice. When the pre-crack does not disrupt these SSS zones, the damage path also localises at the pinned boundaries where these states localise. The damage initiation is shifted back to the pre-crack edge by increasing the pre-crack length or reducing the SR. Notwithstanding the dominance of the SSS, the FM propagating from the pre-crack tip continue to influence the bending stresses here and into the bulk. As such, the findings for roller constraints remain relevant for pinned boundary conditions.

Over a range of SR, we provide insight into the dominant behaviour and failure of topologically distinct kagome lattices. At large SR we are able to accurately map the stress and damage to the underlying microstructure, advancing our understanding of the mechanisms of damage in these lattice-based metamaterials. \rev{Although only two boundary conditions are examined, the form of the edge states are independent of these, making the findings applicable to a wider range of boundary conditions.} These findings provide a robust framework that can be leveraged to architect damage paths themselves. Moving towards a continuum description, this provides the necessary physics to inform the fracture process of these metamaterials.

\begin{acknowledgments}
The authors thank UKRI for support under the EPSRC Open Fellowship scheme (Project No. EP/W019450/1).
\end{acknowledgments}

\section*{Author contributions}
L.d.W. and M.A.D. designed the research; M.A.D. supervised and acquired funding; L.d.W. developed the numerical analysis; L.d.W., M.C., and M.A.D. performed research; L.d.W., M.C., and M.A.D. analyzed data; L.d.W. wrote original draft; and L.d.W., M.C., and M.A.D. wrote and edited the paper.

\appendix

\section{Nomenclature}

\rev{\begin{tabbing}
\hspace{2cm} \= \kill
\textbf{Symbols} \\
$A$ \> Cross sectional area of bonds \\
$A_{edge}$ \> Projected area of the edge $A_{edge} = bL$ \\
$\mathbf{C}$ \> Compatibility matrix \\
$C_\mathrm{s}$ \> Torsional spring stiffness \\
$E$ \> Strain energy \\
$E_\mathrm{m}$ \> Elastic modulus \\
$F$ \> Applied force \\
$F_\mathrm{tol}$ \> Tolerance for convergence of FIRE 2.0 algorithm \\
$H$ \> Initial height of the lattice \\
$I$ \> Second moment of area \\
$K_\mathrm{i}$ \> Axial spring stiffness \\
$L$ \> Length of the domain \\
$N$ \> Number of sites (nodes) in the lattice \\
$N_\mathrm{c}$ \> Number of constraints in the lattice \\
$N_\mathrm{0}$ \> Number of floppy modes (FM) \\
$N_\mathrm{s}$ \> Number of states of self-stress (SSS) \\
$P_\mathrm{T}$ \> Topological polarisation vector \\
$\mathbf{Q}$ \> Equilibrium matrix \\
SR \> Slenderness ratio, SR $= |a_1|/t$ \\
$\textbf{a}_1$, $\textbf{a}_2$, $\textbf{a}_3$  \> Primitive lattice vectors \\
$b$ \> Depth of bonds \\
$\mathbf{d}$ \> Dimensions \\
$\mathbf{e}$ \> Bond elongation vector \\
$\mathbf{f}$ \> Site force vector \\
$l_\mathrm{i}$, $l_\mathrm{i}^\mathrm{0}$ \> Current and original lengths of axial springs \\
$l_\mathrm{\theta}$ \> Length of the boundary within which the SSS at the long wavelength limit may develop \\
$\textbf{p}_1$, $\textbf{p}_2$, $\textbf{p}_3$  \> Vectors normal to the bonds \\
$q_\mathrm{1}, q_\mathrm{2}$ \> Wavenumber components in the \textit{1} and \textit{2} directions \\
$\textbf{r}_1^\mathrm{0}$, $\textbf{r}_2^\mathrm{0}$, $\textbf{r}_3^\mathrm{0}$  \> Basis vectors for the sites of the RK \\
$\mathbf{t}$ \> Bond tension vector \\
$t$ \> Thickness of bonds \\
$\mathbf{u}$ \> Site displacement vector \\
$x_\mathrm{\mu}$ \> Distortion parameter affecting the $\mathrm{\mu}$-th filament \\
$y_\mathrm{j}$ \> Site degrees of freedom\\
$z$ \> Coordination number of a lattice \\
$z_\mathrm{s}$ \> Triangle scaling parameter in kagome lattice \\
$\alpha_\mathrm{GH}$ \> Expected angle between bonds $i$ and $l$ under deformation following the GH mode \\
$\alpha_\mathrm{s}$, $\alpha_\mathrm{0}^\mathrm{s}$ \> Current and original angles in torsional springs \\
$\delta$ \> Applied displacement \\
$\epsilon_\mathrm{n}$ \> Nominal strain, $\epsilon_\mathrm{n} = \delta/H$ \\
$\kappa_\mathrm{1}, \kappa_\mathrm{2}$ \> Imaginary part of the wave numbers in the \textit{1} and \textit{2} directions \\
$\phi$ \> Amplitude of the floppy mode in the bulk \\
$\sigma_\mathrm{A}$, $\sigma_\mathrm{B}$ \> Axial and bending stresses in lattice elements \\
$\tilde{\sigma}_\mathrm{A}$, $\tilde{\sigma}_\mathrm{B}$ \> Normalised axial and bending stresses in lattice elements, $\tilde{\sigma}_\mathrm{A}$, $\tilde{\sigma}_\mathrm{B} = \sigma_\mathrm{A}/\sigma_\mathrm{A}^\mathrm{u}$, $\sigma_\mathrm{B}/\sigma_\mathrm{B}^\mathrm{u}$\\
$\sigma_\mathrm{A}^\mathrm{u}$, $\sigma_\mathrm{B}^\mathrm{u}$ \> Material failure stresses in axial and bending\\
$\sigma_n$ \> Normalized applied stress \\
$\theta$ \> Propagation angle of floppy modes \\
\\
\textbf{Abbreviations} \\
AR \> Aspect Ratio \\
D-FM \> Domain wall of Floppy Modes \\
D-SSS \> Domain wall of States of Self-Stress \\
DW \> Domain wall\\
FM \> Floppy Modes \\
GH Mode \> Guest-Hutchinson Mode \\
LP \> Left Polarised \\
LPU \> Left-Up Polarised \\
LTK \> Left Twisted kagome \\
NN \> Nearest Neighbour \\
NNN \> Next Nearest Neighbour \\
RK \> Regular kagome lattice \\
RP \> Right Polarised \\
RDP \> Right-Down Polarised \\
RTK \> Right Twisted kagome \\
SSS \> States of Self-Stress \\
TK  \> Twisted kagome \\
\end{tabbing}}

\section{\rev{Lattice definitions}}\label{LatticeDeffinition}

\begin{figure*} 
	\centering
	\includegraphics{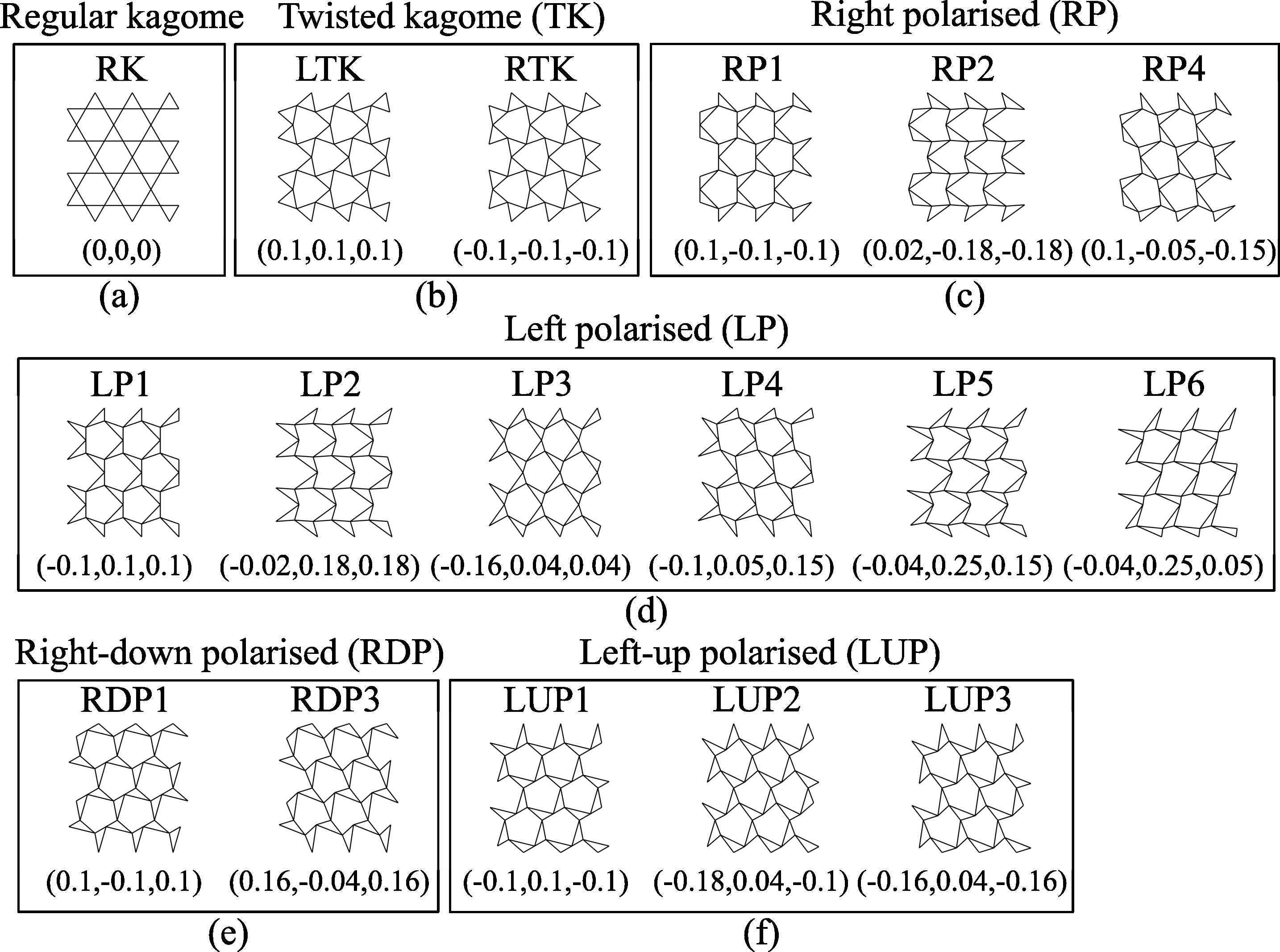} %
	\caption{\rev{Lattice definition with values of the parametrisation $\{x_\mathrm{1},x_\mathrm{2},x_\mathrm{3}\}$ for: (a) Regular kagome (RK); (b) Twisted kagome (TK); (c) Right polarised (RP); (d) Left polarised (LP); (e) Right-down polarised (RDP); and, Left-up polarised (LP)}}
    \label{fig:S1a} 
\end{figure*}

\rev{A summary of the lattices simulated within this study are provided in Fig. \ref{fig:S1a}.}

\section{Method}\label{Method}

In this section we provide further detail about the numerical simulator built to model the damage process. Information on the minimisation algorithm, failure criteria and inputs can be found in the main text.  

\subsection{Discretisation}\label{ADiscr}
Axial spring $i$ connect adjacent sites $j$ and $k$; and, torsional spring $s$ connects adjacent bonds $i$ and $l$ (Fig. \ref{fig:2}a). When more than two bonds are connected at a site, torsional springs are added between all adjacent bonds (Fig. \ref{fig:2}b). In all simulations, we discretise each bond into three elements. \rev{This level of discretization cannot resolve modes beyond the second buckling mode. However, given the quasi-static loading conditions and the constraints applied in our simulations, higher order deformation modes are not expected to occur.}

\subsection{Spring stiffness}\label{ASpringStiff}

The axial and bending stiffness of the two springs can be found as follows:
\begin{equation}\label{Eq:S1}
    K_\mathrm{i} = \frac{E_\mathrm{m}A}{l_\mathrm{i}^\mathrm{0}}
\end{equation}
\begin{equation}\label{Eq:S2}
    C_\mathrm{s} = \frac{E_\mathrm{m}I}{\min(l_\mathrm{i}^\mathrm{0},l_\mathrm{l}^\mathrm{0})}
\end{equation}
Where $E_\mathrm{m}$ is the elastic modulus, $A$ is the cross sectional area, $I$ is the second moment of area and $l_\mathrm{i/l}^\mathrm{0}$ are the initial length of the elements. The length of bond $i$ and $l$, which are connected by torsional spring $s$ may not be equivalent. In this case, the minimum length of the two bonds was used to determine $C_\mathrm{s}$. Little difference was observed when taking the average length or maximum length.

\subsection{Stress in bonds}\label{AStress}

The force at each site is minimised to find the equilibrium configuration of the lattice within each iteration. Once this is complete, the stress in the cross section due to axial elongation ($\sigma_\mathrm{A}$) and bending ($\sigma_\mathrm{B}$) is calculated at both sites $j$ and $k$ of each bond $i$ as follows:
\begin{equation}\label{Eq:S3}
    \sigma_\mathrm{A,i} = \frac{K_\mathrm{i}}{A} (l_\mathrm{i} - l_\mathrm{i}^\mathrm{0})
\end{equation}
\begin{equation}\label{Eq:S4}
    \sigma_\mathrm{B,i} = \frac{t}{2IA} \sum_\mathrm{s=1}^{n_\mathrm{r}} C_\mathrm{s}(\alpha_\mathrm{s}-\alpha_\mathrm{s}^\mathrm{0}) 
\end{equation}
The moment ($\sum_\mathrm{s=1}^{n_\mathrm{r}} C_\mathrm{s}(\alpha_\mathrm{s}-\alpha_\mathrm{s}^\mathrm{0})$) at each end of the bond has contributions from the $n_\mathrm{r}$ torsional springs that are connected to bond $i$ at that end.

\subsection{Algorithm}\label{AAlgo}

Multiple iterations may occur within each step to simulate the damage process; however, only one site can be split within each iteration. When a site is split within an iteration, the forces at the sites are minimised for the updated geometry in the following iteration. This captures the redistribution of the stress associated with the fracture. Within this iteration, we then check if any elements are above the threshold and the process repeats until no element is above the threshold. If there are no bonds above the threshold, the step is completed, and the next step is taken from this position. 

If one element is above the threshold, the associated site is split and the forces at the sites are minimised for the updated geometry. Within this iteration, we then check to see if any elements are above the threshold, and if no element is above this threshold, the step is completed. If one element is above the threshold the same process is followed again. If there is more than one element above the threshold, rather than splitting all associated sites, only the element with the largest stress has its site split. The force minimisation is then applied and the process is followed again. The iterations within the steps continue until no element is above the threshold value. 

If more than one element is above the threshold in the initial step, the step size is halved, and the initial step is retaken. If, in the subsequent iteration, multiple elements are still above the threshold, only the site associated with the maximum element stress is split, and the same process as outlined above is followed.

\section{Evanescent Mode Visualizations}\label{Evanescent_Mode_Appendix}

In this section, we explain the methods used to visualize the form of evanescent modes associated with bulk discontinuities. It is important to mention that although the solutions we visualize given a surface disturbance are exact, the (functional) form of this disturbance has to be prescribed following certain assumptions. The visualizations are thus an instructive means of designing the lattices. For instance, in the case of FM on the crack surface of LP lattices, we use the qualitative assumption of a strongly localised disturbance on the crack tip, which can be motivated through fracture mechanics on micropolar media \citep{BerkacheKamel2022Meot}, as well as our simulation results (see Fig.2i). For RP lattices, which do not possess a significant spike, we have instead used a sigmoid prescription, which aims to emulate the relaxation of the lattice on the crack-edge, and the fact that the disturbance approaches a uniform actuation in the bulk. Finally, although our examples are focused on FM visualisations, note that the same method can be applied for SSS.

\subsection{Bulk-Boundary Correspondence}
In the limit of linear elasticity, the existence of uniform strain, zero energy (\textit{e.g.} GH) modes imply the existence of spatially modulated versions of the same strain that also cost no energy \citep{rocklin2017transformable}. For idealized, zero bending stiffness kagome lattices, the zero energy distortion on the unit cell can be captured by a single parameter $\phi$ \citep{SunKai2012Sper, NassarHussein2020MeAc}, which, when allowed to vary spatially defines a type of micro-deformation field $\phi(\boldsymbol{x})$. In the linear elastic limit, geometric compatibility has to be enforced, which constrains the functional form for $\phi(\boldsymbol{x})$, effectively reducing the degrees of freedom of the function from two independent coordinates $\phi \equiv \phi(x,y)$ into a combination of the two $\phi \equiv \phi(x + \lambda y)$.

This phenomenon, known as \textit{bulk-boundary correspondence}, is also apparent in the full-field (non-linear) solution, where it can be understood as a restatement of the zero energy requirement. In Fourier space we can relate the orthogonal components of the wavenumbers describing the (floppy) distortion of the lattice, $q_\mathrm{1}$ and $q_\mathrm{2}$ through this constraint:
\begin{equation} \label{Eq:A5}
    E(q_\mathrm{1}, q_\mathrm{2}) = \det\mathbf{C}(q_\textrm{1},q_\textrm{2}) = 0.
\end{equation}
allowing us to write, \textit{e.g.} $q_\mathrm{2} \equiv G(q_\mathrm{1})$, prescribing one wavenumber component in terms of the other. Given that $C(q_\mathrm{1}, q_\mathrm{2})$ contains all information about the lattice, this model does not make any homogenising assumptions - $q_\mathrm{2} = G(q_\mathrm{1})$ is exact, $\forall q_\mathrm{1} \in \left[-\pi/a, \pi/a \right]$. Note that, as $\mathbf{C} = \mathbf{Q}^\dagger$, the $q_\mathrm{2}$ component of the SSS will be the complex conjugate of the FM resulting in $\kappa^\mathrm{SSS} = -\kappa^\mathrm{FM}$ and $\theta^\mathrm{SSS} = \theta^\mathrm{FM} = \theta$---although we focus on FM, the extension to the discussion of SSS is straightforward. Now, using this result, consider the deformation field $\phi(\boldsymbol{x})$, written as an inverse Fourier Transform:
\begin{align}
    \phi(x, y) &= \int_{-\infty}^\infty dq_\mathrm{1} e^{i q_\mathrm{1} x} \int_{-\infty}^\infty dq_\mathrm{2} e^{i q_\mathrm{2} y}  \ \Tilde{\phi}(q_\mathrm{1}, q_\mathrm{2}) \delta(q_\mathrm{2} - G(q_\mathrm{1})) \nonumber \\
    &= \int_{-\infty}^\infty dq_\mathrm{1} \Tilde{\phi}(q_\mathrm{1}) e^{i q_\mathrm{1} x} e^{i G(q_\mathrm{1}) y}, \label{Eq:S6}
\end{align}
where we have used the Dirac delta $\delta(q_\mathrm{2} - G(q_\mathrm{1}))$, to signify that the value of $q_\mathrm{2}$ is fully determined by $q_\mathrm{1}$, and have performed the $q_\mathrm{2}$ integral using the sifting property.

Suppose that we know the distortion field on a linear region $y = y_0$, i.e. $\phi(x, y_0) = F(x)$, where $F(x)$ is some known function. The bulk-edge correspondence, restated as \eqref{Eq:S6}, allows us to reconstruct the full field solution by first retrieving $\Tilde{\phi}(q_\mathrm{1})$ using the known information about the field on $y=y_0$, and then computing the inverse FT.

\begin{figure*} 
	\centering
	\includegraphics{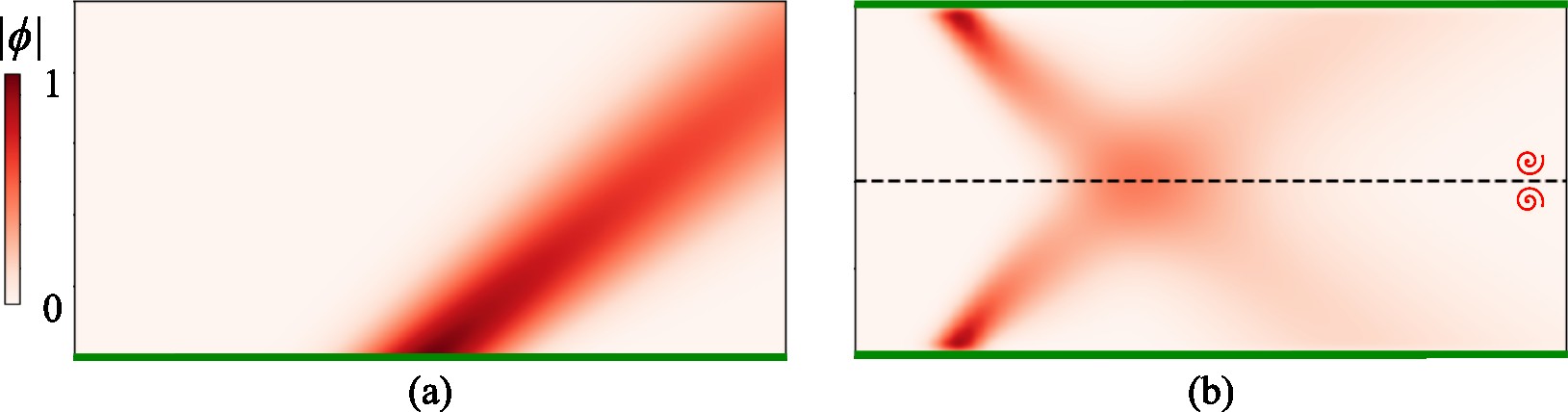} %
	\caption{Continuum surface mode visualizations. Demonstrations of disturbances with, (a) open and (b) roller boundary conditions, for LP1. These continuum plots depict the modulus $|\phi|$, of the otherwise complex FM field. In all cases, green solid lines represent free edges, with Gaussian distortions of varying widths. The dashed black line in (b) represents a boundary with rollers, delineating the top half domain, which is virtual and contains the mirror image of the field in the lower half - note the two vortexes highlighting the mirroring.}
    \label{fig:S1} 
\end{figure*}

In a discrete domain comprised of $N \times N$ equally spaced sampling points in a square lattice configuration, we can write \eqref{Eq:S6} as an inverse Discrete Fourier Transform.

\begin{equation}
 \phi_{l,m} = \sum_{n=0}^{N-1} \Tilde{\phi}_n e^{i \frac{2 \pi n}{N} l} e^{i G\left(\frac{2 \pi n}{N} \right) m},
\end{equation}
with $l,m$ indexing the $x,y$ locations in the grid, and where we are summing over the wavenumbers $q_\mathrm{1} \equiv 2\pi n/N$; $G(q_\mathrm{1})$ being the geometry dependent function we retrieve through the zero energy constraint \eqref{Eq:A5}. Then, the Fourier components $\Tilde{\phi}_n$ can be found through a prescription of the field on N points e.g. on the horizontal line defined by $m = m_0$,
\begin{equation}
    \Tilde{\phi}_n = \sum_{l=0}^{N-1} \phi_{l,m_0} \ e^{-i \frac{2 \pi n}{N} l}.
\end{equation}
The above algorithm is sufficient to produce a full numerical ``solution'', given a certain known disturbance $\phi_{l,m_0}$ in a domain.

\subsection{FM Prescriptions}
For prescribing the FM associated with the crack, we use the centre-line of the domain where the crack surface lies and use simple functions that produce qualitatively similar disturbances to the ones observed in Fig.2i. For example, to emulate the highly localised spike in LP and LTK lattices, we use a narrow Gaussian disturbance of unit height:
\begin{equation}
    \phi_{\mathrm{l}, \mathrm{m}_\mathrm{0}} = e^{(l - l_\mathrm{c})^2/2\sigma^2},
\end{equation}
where $l_c$ is the location of the crack tip, with a width of $\sigma$, set to one unit cell length. In the case of RP, we choose a sigmoid disturbance centered on the crack tip. In particular, we use:
\begin{equation}
    \phi_{l,m_0} = \sigma(l - l_\mathrm{c}) = \frac{1}{1+e^{-(l - l_\mathrm{c})}}.
\end{equation}
This choice derives from the dominant contribution to the field which exhibits a contrast between the uniform actuation in the bulk and the relaxation of the lattice along the crack edge. The choice of smooth functions is deliberate, as discontinuous functions, such as the \textit{Heaviside} or \textit{Kronecker delta}, give rise to ringing artifacts in the bulk. Of course, multiple disturbances can be superimposed; however, here, we only visualise the dominant actuation to capture the most important elements of the bending stress distribution in the bulk.

\subsection{Boundary Conditions}

A FM, in general, distorts the lattice in a way that does not respect the Boundary Conditions (BC). An equivalent way of viewing the continuation of the disturbance around the boundary constraint is to consider a secondary disturbance $\phi_b(x,y)$ localized on the boundary, which, when superimposed with the primary disturbance localized on the crack tip $\phi_0(x,y)$, has to give a displacement field that respects the BCs. For the strongly localised crack tip actuation, there are two sets of Dirichlet BCs we use in this work: (1) rollers ($\delta y = 0$) and (2) pinned ($\delta x = 0$, $\delta y = 0$). Although pinning the sites lifts the FM \citep{kane2014topological}, the roller boundary conditions may allow for their existence by a reflection-like propagation that we examine below.

For y-mirror-symmetric unit cells (LP, RP, TK), roller boundary conditions can be satisfied by superimposing two images of the disturbance, mirrored around the boundary. This is because the mirroring achieves the desired $\delta y_b = -\delta y_0$, on the boundary (see Fig. \ref{fig:S1}b). Furthermore, the continuation of the real domain into its virtual counterpart is compatible with the unit cell geometry, as the unit cell is chiral. For achiral unit cells that still belong in the equivalence classes $(\pm, \pm, \pm)$ or $(\mp, \pm, \pm)$, mirroring will cause a discontinuity in the lattice tiling around the boundary. Nonetheless, mirroring would still satisfy the BC, with the extension of the FM from one domain to the other refracting the disturbance. Numerically, the FM can be constructed together with its reflection by obtaining the primary amplitude on the boundary $\phi_0(x, y_b)$, and then using this as the new prescription for the secondary mode $\phi_b(x,y)$.

\begin{figure*} 
	\centering
	\includegraphics{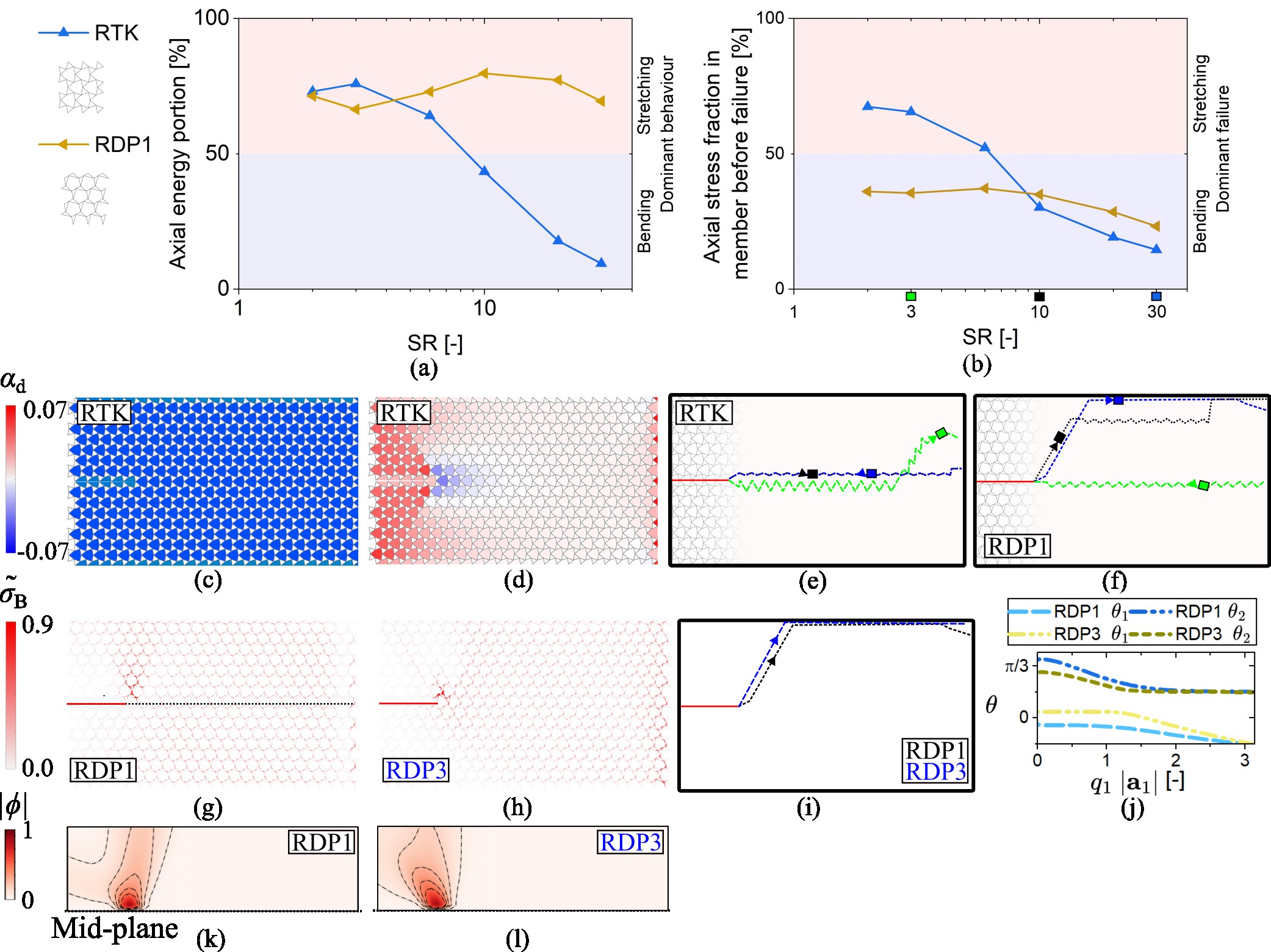} %
	\caption{\rev{Dominant behaviour and failure of additional lattices. Snapshot of: (a) axial energy portion in the entire lattice; and, (b) axial stress fraction in the element that fails preceding the first failure event. Comparison of the internal angle of a unit cell from the simulations ($\alpha_\mathrm{s}$) with that required by the GH mode ($\alpha_\mathrm{GH}$) to displace the boundary the equivalent distance (\emph{i.e.} $\alpha_\mathrm{d} = \alpha_\mathrm{s}-\alpha_\mathrm{GH}$) for RTK at: (c) SR = 3; and, (d) 30. The difference in angle is visualised through the colour directly to the right of each unit cell. Damage path for: (e) RTK; and, (f) RDP1. The damage process is visualised with a line traversing the domain. The line connects separated nodes, where fracture occured, within the lattice. For clarity, we do not show the micro-structure throughout the domain. The damage process for SR = 3, 10, 30 is plotted with green, black and blue lines and associated rectangular markers respectively (refer to (b)). Arrows indicate the direction of damage. Normalised maximum bending stress in cross section $\sigma_\mathrm{\textbf{Bend}} = |\sigma_\mathrm{B}/\sigma_\mathrm{B}^\mathrm{u}|$ and visualisation of FM through the magnitude of the total modulation ($|\phi|$) of the wave spectrum for: (g, k) RDP1 (0.1,-0.1,0.1); and, (h, l) RDP3 (0.16,-0.04,0.16) at SR = 30. (i) Damage path for RDP1 and RDP3 at SR = 30. (j) $\theta$ spectrum for RDP1 and RDP3.}}
	\label{figS2a} 
\end{figure*}

For lattices in the $(\pm, \mp, \pm)$ or $(\pm, \pm, \mp)$ equivalence classes and initial geometries with a horizontal crack, like the ones considered in this work, both modes localize on the same horizontal edge, and the disturbance cannot be canceled on the boundary by a mode localizing there. It is not directly clear what combination of modes localizing on the remaining edges is induced in the lattice, so as to respect the boundary conditions. As a result, we limit our visualizations to the case of open boundary conditions (e.g. Fig. \ref{fig:S1}a) for these modes.

\subsection{Dominant SSS}\label{PinnedSizeEffects} 
\begin{figure*}
	\centering
	\includegraphics[width=0.97\textwidth]{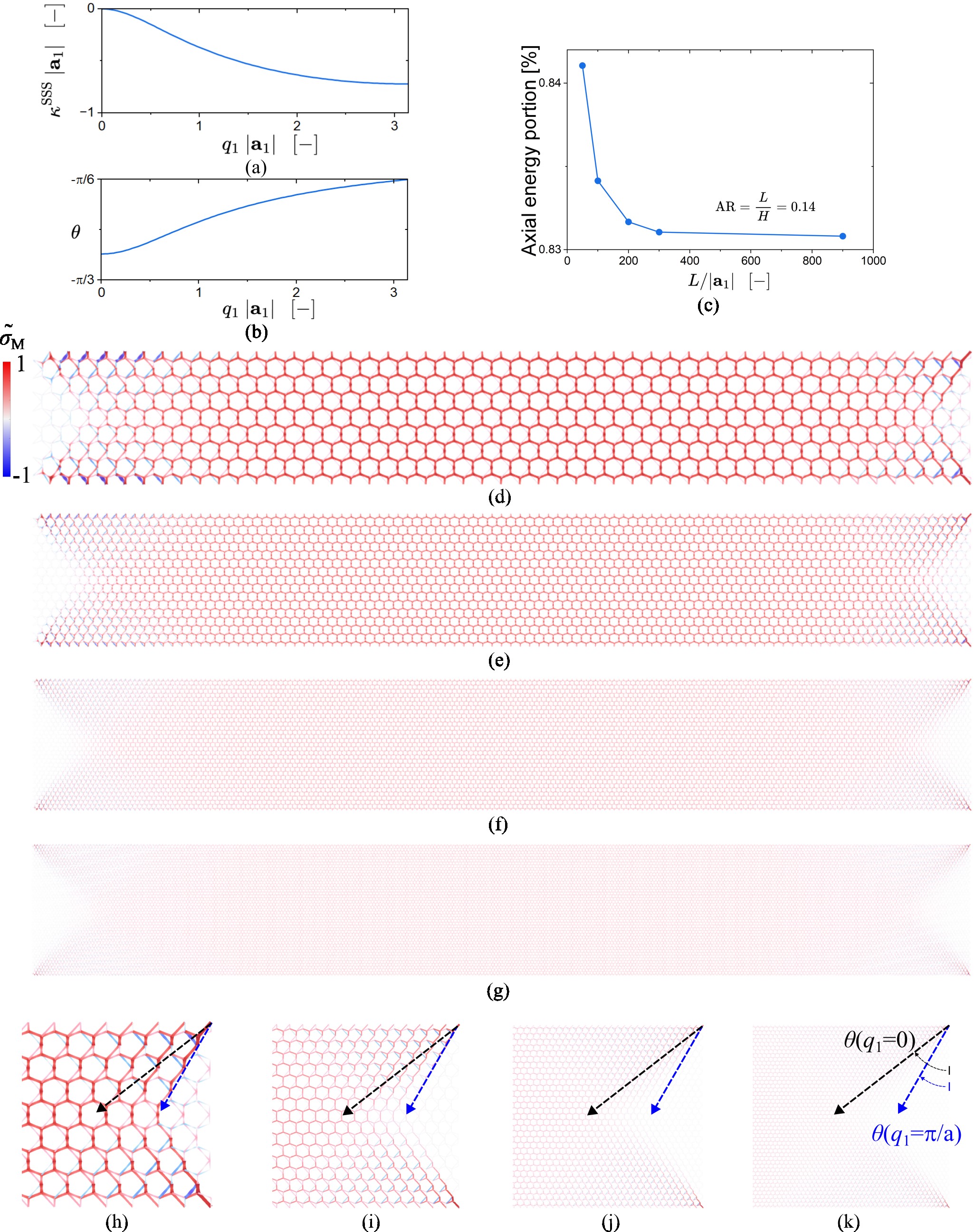} %
	\caption{Size effect study for LP1 with AR = 0.14. (a) $\kappa^\mathrm{SSS}$; (b) $\theta$; (\textbf{C}) axial energy portion vs length of domain at $\epsilon_{\mathrm{n}} = 0.01$; and, total normalised stress profile ($\sigma_\mathrm{M} = \sigma_\mathrm{A}/\sigma_\mathrm{A}^\mathrm{u} \pm |\sigma_\mathrm{B}/\sigma_\mathrm{B}^\mathrm{u}|$) for lattice with: (\textbf{D, H}) \textit{n} = 8, \textit{m} = 50; (\textbf{E, I}) \textit{n} = 16, \textit{m} = 100; (\textbf{F, J}) \textit{n} = 32, \textit{m} = 200; and, (\textbf{G, K}) \textit{n} = 48, \textit{m} = 300.}
	\label{figS3a}
\end{figure*}

\section{Behaviour of additional topologically distinct lattices}\label{AdditionalLattices}

In this section, we summarise the results for two additional lattices with $\mathrm{sgn}(x_\mathrm{1}, x_\mathrm{2}, x_\mathrm{3})$ of (-,-,-) and (+,-,+). (-,-,-) corresponds to RTK which is indistinguishable topologically from LTK.  In contrast, (+,-,+), related to RDP, is a separate equivalence class not reported in the main text. 

\subsection{Dominant lattice behaviour}

The dominant behaviour for RTK (-0.1, -0.1, -0.1) follows an identical trend to LTK, transitioning from stretching to bending as the SR increases (Fig. \ref{figS2a}a). The GH mode for RTK maps onto the applied boundary constraints, and therefore, a bending deformation mode is favoured at high SR (Fig. \ref{figS2a}d). While RTK is topologically equivalent to LTK, the FM propagate to the left rather than the right. This can be observed in Fig. \ref{figS2a}c where large differences to the idealised GH modes are observed diagonally to the left of the crack tip. As SR decreases, the deformations diverge from this mechanism, preferentially stretching (Fig. \ref{figS2a}c). In contrast, RDP1 with (0.1, -0.1, 0.1) is always stretching-dominated (Fig. \ref{figS2a}a) as its GH mode does not map onto the boundary conditions applied.

\subsection{Dominant lattice failure}

RTK dominant failure switches from bending to stretching as the SR reduces (Fig. \ref{figS2a}b), with the damage path consistently deviating from the mid-plane (Fig. \ref{figS2a}e). At SR = 10 and 30, where the fracture is dominated by bending, the damage paths deviate slightly above the pre-crack before propagating with small oscillations towards the right edge. At SR = 3, the failure becomes stretching-dominated, with larger oscillations in the damage path occurring as it propagates toward the right edge. In contrast, RDP1 fracture is always dominated by bending (Fig. \ref{figS2a}b). At SR = 30, the bending stresses and damage path (Fig. \ref{figS2a}g) align with the FM localising at the crack tip (Fig. \ref{figS2a}\rev{k}); the damage path can therefore be adjusted through their form. This is the case for RDP3, where $\theta_\mathrm{1}$ is initially positive (Fig. \ref{figS2a}\rev{k}), and the fracture initiates further towards the left of the domain than RDP1. 

\section{Additional constraints}\label{AdditionalConstraints}

When pinned boundary conditions are added, FM are removed and/or SSS are added to the system. Here we provide further results for lattices with these boundary constraints to highlight how they influence the behaviour and damage. 

The SSS spectrum for LP1 is provided in Fig. \ref{figS3a}a,b. In the main text,  $l_\mathrm{\theta}$ is introduced as a variable that defines the number of unit cells over which the SSS spread. Unlike the FM, which localise at the crack tip for LP1, $l_\mathrm{\theta}$ may span many unit cells. Broadening this stress field results in the dominant wavenumber tending towards the long-wavelength limit, which does not decay \emph{i.e.} propagating infinitely into the bulk. Therefore, provided $l_\mathrm{\theta}$ is sufficiently large, we assume that the dominant behaviour for LP1 will not be greatly affected by size effects. To probe this phenomena we simulate LP1 with increasing lattice length $L$, whilst keeping the aspect ratio (AR = $H/L$) constant at 0.14. This AR was chosen so that $L$ is more than 5 times that required for the SSS at the long-wavelength limit to develop (\emph{i.e.} $L > 5 H \tan{\left[\theta(q=0)\right]}$). The ability to develop refers to the SSS at the long-wavelength limit meeting a restrained edge rather than a free edge. Comsol was used to study the change in dominant behaviour with increasing lattice size where \textit{n} and \textit{m} represents the number of unit cells along \textit{H} and \textit{L} respectively. Identical material properties as those used in the simulator were implemented and the lattice was discretised with a maximum beam element size of 0.5mm. 

Fig. \ref{figS3a}d-k provides the plots of the normalised stress within the elements for four separate domain sizes. The snapshots are taken at a nominal strain of 0.01. At the right edge, close to the restrained edge, the stress appears to align along the short-wavelength limit (Fig. \ref{figS3a}a,b). However, further into the bulk the stress associated with this direction decays, resulting in the stresses tending towards the long-wavelength limit. This trend is not observed for Fig. \ref{figS3a}d,h as the domain size is not large enough for this SSS to decay appreciably. 

Fig. \ref{figS3a}c provides the axial energy portion within the domain at this nominal strain. Only a small difference is observed across the range of lengths simulated, with the axial energy portion increasing as $L$ reduces. This small difference is attributed to the proportion of the lattice subjected to the short-wavelength limit SSS. As the domain size increases, this has a diminishing effect, and the axial energy converges to a value below this. Regardless, for this AR and $L$ range simulated, LP1 response is dominated by the SSS at the long-wavelength limit. At larger AR or lattices where the form of the SSS reduces $l_\mathrm{\theta}$, it is expected that size effects become more appreciable. However, no studies have been undertaken at this stage to investigate this phenomena.

\subsection{Failure evolution for LP2 lattice with larger length}\label{LP2Larger}

\begin{figure}
	\centering
	\includegraphics[scale=0.90]{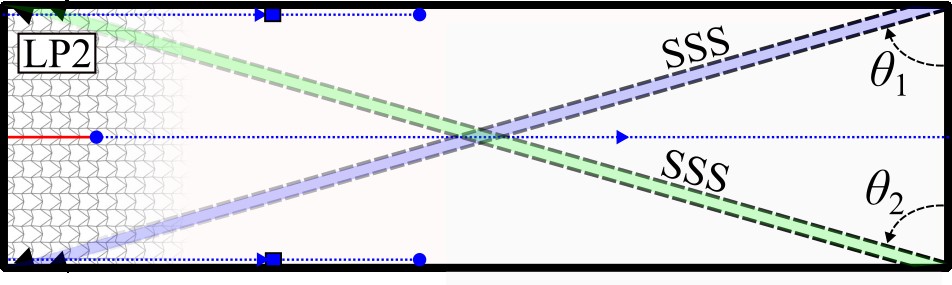} %
	\caption{Damage propagation for LP2 with n = 16, m = 50 at SR = 30. The simulations for this lattice is included in Movie S5.}
	\label{figS4a} 
\end{figure}

In the main text, we provide results for LP2 with $m = 24$. For this domain size the SSS at the long-wavelength limit does not develop. Fig. \ref{figS4a} shows the failure path for LP2 with $m = 50$, a domain length that allows this SSS to develop. Instead of the damage horizontally propagating from the crack tip, failure initiates at the top and bottom left edges, where the SSS end, subsequently propagating horizontally towards the right edge. 

\begin{figure*} 
	\centering
	\includegraphics{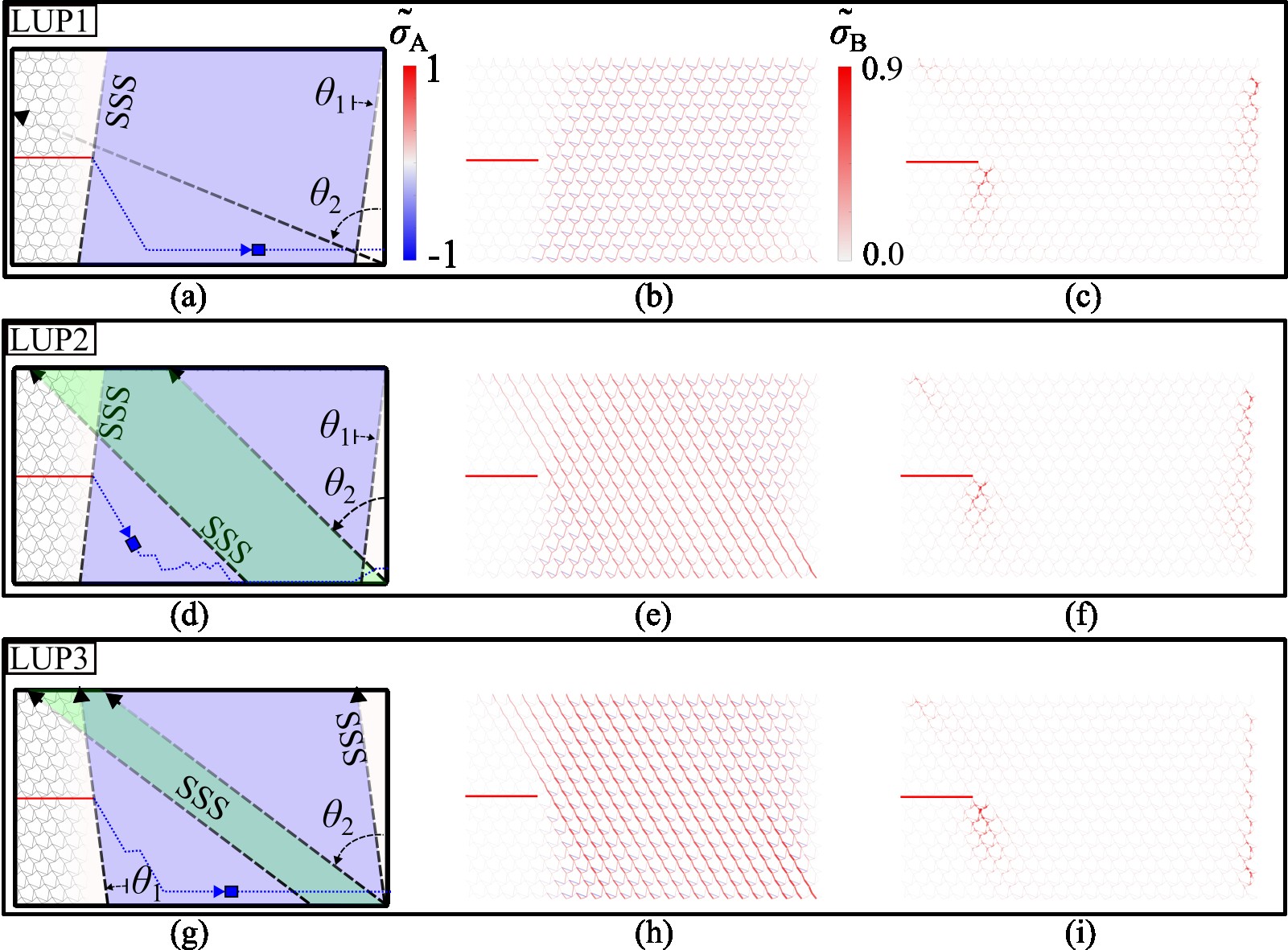} %
	\caption{Stress development in LUP lattices with pinned boundary conditions. Failure path, normalised axial and normalised bending stress for: (a-c) LUP1; (d-f) LUP2; and, (g-i) LUP3 respectively at SR = 30 with pinned boundary conditions. In (a), (d) and (g), we also highlight the zones where the SSS at the long-wavelength limit may develop with arrows on either side of these zones. If the SSS can't develop we only plot the direction of the SSS with a single arrow. The simulations for these lattices are included in Movie S5.}
	\label{figS5a} 
\end{figure*}

\subsection{Stress profile in lattices from ($-$,$+$,$-$) equivalence class}\label{LUPPinned} 

Here we provide the results for LUP1-3, all from the ($-$,$+$,$-$) equivalence class, to further highlight the influence of the SSS form on the stress that develops within the domain and around the crack tip. The $\theta$ spectrum for these lattices is provided within the main text. The stresses within the lattices map onto the SSS at the long-wavelength limit in all cases. For LUP1, only $\theta_\mathrm{1}$ can develop within the domain (Fig. \ref{figS5a}(a)), and in Fig. \ref{figS5a}(b) we see that the axial stress is within these zones. As SSS at $\theta_\mathrm{1}$ dominate the response, the bending stresses that are activated at the crack tip also follow the FM associated with $\theta_\mathrm{1}$ (Fig. \ref{figS5a}(c)). In LUP2 the microstructure of the lattice is adjusted to reduce $\theta_\mathrm{2}$ at the long-wavelength limit whilst keeping the $\theta_\mathrm{1}$ spectrum similar to LUP1 (Fig. \ref{figS5a}(d)). Compared with LUP1, additional axial stresses aligning with $\theta_\mathrm{2}$ appear (Fig. \ref{figS5a}(e)). However, as the pre-crack only intersects the SSS zone associated with $\theta_\mathrm{1}$, the bending stresses around the crack tip for LUP2 remain similar (Fig. \ref{figS5a}(f)). The initial damage path also matches LUP1, however as the damage evolves, the path for LUP2 (Fig. \ref{figS5a}(d)) deviates from LUP1 (Fig. \ref{figS5a}(a)). This may be due to the interaction with the $\theta_\mathrm{2}$ SSS, although this has not been investigated further here. $\theta_\mathrm{1}$ becomes positive at the long-wavelength limit for LUP3 (Fig. \ref{figS5a}(g)). The axial stresses that develop within the domain also follow this form (Fig. \ref{figS5a}(h)) and the bending stress about the crack tip shift in line with these changes (Fig. \ref{figS5a}(i)).

\section{Achiral lattices from the ($\mp$,$\pm$,$\pm$) equivalence classes}\label{Achirallattices}

All lattices reported in the main text from the ($\mp$,$\pm$,$\pm$) equivalence classes had FM and SSS with forms that where mirror symmetric wrt. the plane of the crack. Here, we provide additional simulations for lattices with asymmetric forms for these mode, \emph{i.e.} achiral unit cell geometries. In particular we choose to visualise: LP4 (0.1, -0.05, -0.15); RP4 (-0.1, 0.05, 0.15); LP5 (0.04, -0.25, -0.15); and, LP6 (0.04, -0.25, -0.05). All simulations are for SR = 30. 
 
\begin{figure*}
	\centering
	\includegraphics{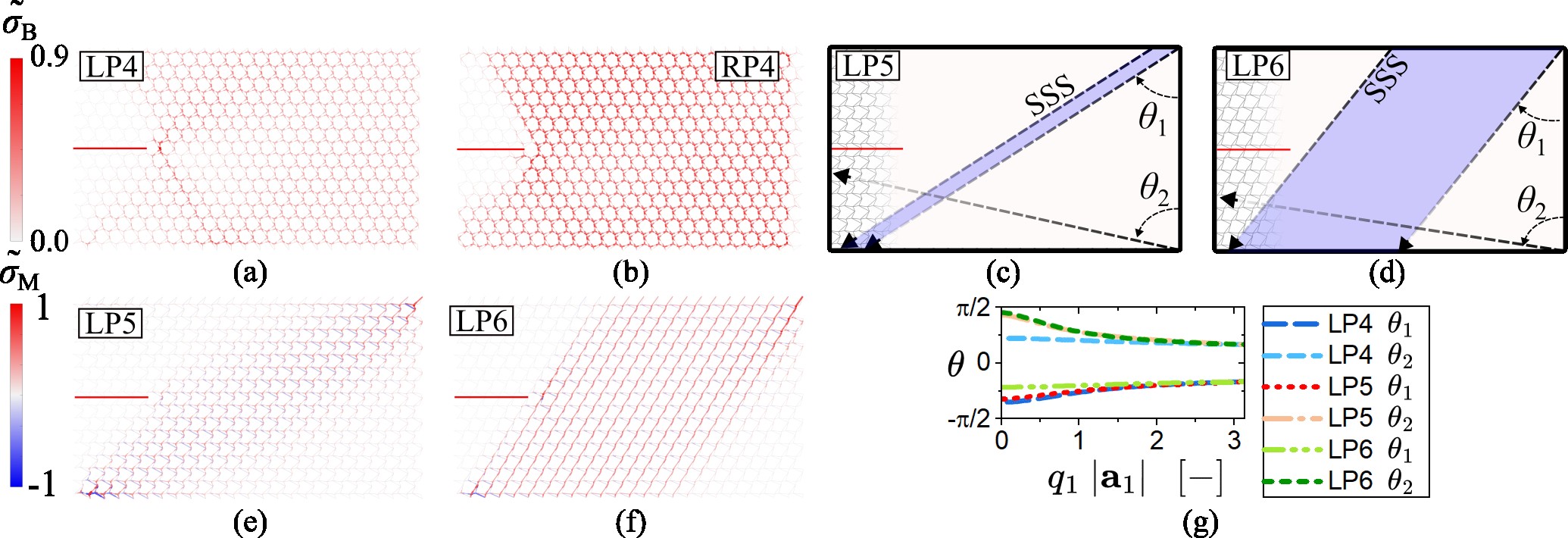} %
	\caption{\rev{LP and RP lattice with achiral FM/SSS forms. Bending stress in: (a) LP4; and, (b) RP4. SSS development at the long-wavelength limit and normalised stress in domain for: (c, e) LP5; and, (d, f) LP5. All simulation results are for SR = 30 with pinned boundary conditions. (g) $\theta$ spectrum for LP4, LP5 and LP6.}}
	\label{figS6a} 
\end{figure*}

While at the short-wavelength limit $|\theta_\mathrm{1}| = |\theta_\mathrm{2}|$, at the long-wavelength limit $|\theta_\mathrm{1}| < |\theta_\mathrm{2}|$ for LP4 and RP4 (Fig. \ref{figS6a}(g)). The bending stress follows this trend, where below (above) the crack tip they spread across the narrower (broader) $\theta_\mathrm{1}$ ($\theta_\mathrm{2}$) spectrum for LP4 (Fig. \ref{figS6a}(a)). A similar trend is observed for RP4, where there is a larger area with lower bending stresses above the crack when compared to below (Fig. \ref{figS6a}(a,b)). 

When pinned boundary conditions are enforced, SSS develop within the lattice. In Fig. \ref{figS6a}(c,d), we show the SSS at the long-wavelength limit for LP5 and LP6 respectively. In both cases $\theta_\mathrm{2}$ can't develop and $\theta_\mathrm{1}$ dominates. $\theta_\mathrm{1}$ is larger for LP5 (Fig. \ref{figS6a}(g)), resulting in the stress developing within a narrower zone when compared to LP6 (Fig. \ref{figS6a}(e,f)).



\bibliography{apssamp}

\end{document}